\documentclass[manuscript]{acmart}

\AtBeginDocument{%
  \providecommand\BibTeX{{%
    \normalfont B\kern-0.5em{\scshape i\kern-0.25em b}\kern-0.8em\TeX}}}

\copyrightyear{2025}
\acmYear{2025}
\setcopyright{rightsretained}
\acmDOI{10.1145/3586183.XXXXX}
\acmISBN{979-8-4007-0132-0/23/XX}

\usepackage{color}
\usepackage{soul}    % highlighting
\usepackage[capitalize,nameinlink]{cleveref}

\graphicspath{ {figures/} }
 
\usepackage{gensymb} % degree symbol

\usepackage{enumitem} % remove indent of list items

\usepackage{xspace} % for putting space when necessary

\newcommand{\new}[1]{\textcolor{black}{#1}}

\newcommand{\finaledit}[1]{\textcolor{black}{#1}}

\makeatletter
\renewcommand{\paragraph}[1]{%
  \noindent\hspace*{\parindent}% Indent the title and text
  {\bfseries\itshape #1}% Bold title
  \hspace{0.5em}% Space between the title and the text
  \ignorespaces% Remove any extra space after the title
}
\makeatother

\newcommand{\systemName}{\textsc{XR-penter}\xspace}

\begin{document}
 

\title[\systemName: Material-Aware and In Situ Design of Scrap Wood Assemblies]{\systemName: % Enabling %Material-Aware, \\ 
Material-Aware and In Situ Design of Scrap Wood Assemblies}
% Situated Constraints?
%  Improvisational,


\author{Ramya Iyer}
\affiliation{%
  \institution{Georgia Institute of Technology}
  \city{Atlanta}
  \country{USA}}
\email{ramyaiyer@gatech.edu}
\orcid{0009-0008-9643-8029}

\author{Mustafa Doga Dogan}
\affiliation{%
  \institution{Adobe Research}
  \city{Basel}
  \country{Switzerland}}
\email{doga@mit.edu}
\orcid{0000-0003-3983-1955}

\author{Maria Larsson}
\affiliation{%
\institution{The University of Tokyo}
 \city{Tokyo}
\country{Japan}}
\email{ma.ka.larsson@gmail.com}
\orcid{0000-0002-4375-473X}

\author{Takeo Igarashi}
\affiliation{%
\institution{The University of Tokyo}
 \city{Tokyo}
\country{Japan}}
\email{takeo@acm.org}
\orcid{0000-0002-5495-6441}


\renewcommand{\shortauthors}{Iyer, et al.}

\begin{abstract} {
Woodworkers have to navigate multiple considerations when planning a project, including available resources, skill-level, and intended effort. \textit{Do it yourself} (DIY) woodworkers face these challenges most acutely because of tight material constraints and a desire for custom designs tailored to specific spaces. To address these needs, we present \systemName, an extended reality (XR) application that supports in situ, material-aware woodworking for casual makers. Our system enables users to design virtual scrap wood assemblies directly in their workspace, encouraging \textit{sustainable} practices through the use of discarded materials. Users register physical material as virtual \textit{twins}, manipulate these twins into an assembly in XR, and preview cuts needed for fabrication. \finaledit{We conducted a case study and feedback sessions to demonstrate how \systemName supports improvisational workflows in practice, the type of woodworker who would benefit most from our system, and insights on integrating similar spatial and material considerations into future work.}}

\end{abstract}




\begin{teaserfigure}
  \centering
  \includegraphics[width=1\textwidth]{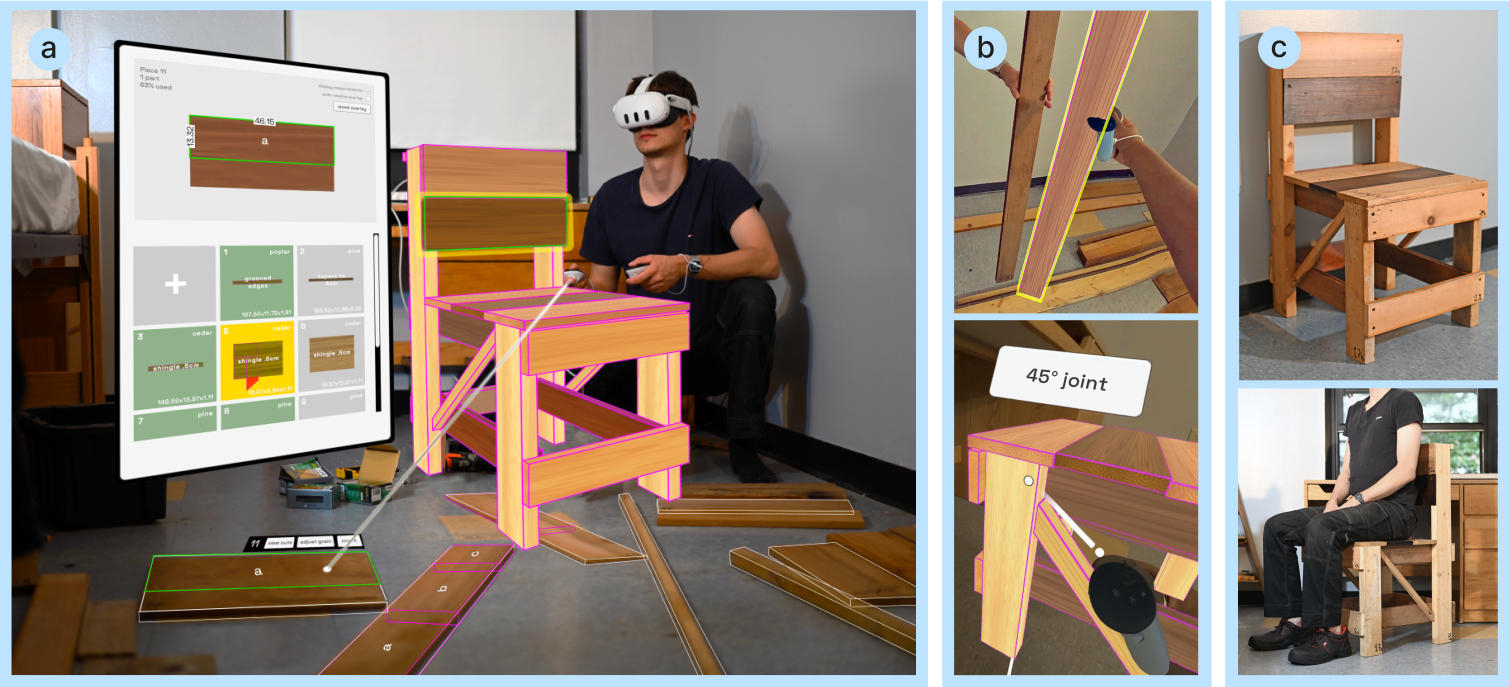}
  \caption{(a)~\systemName enables the user to cut virtual material and assemble it into a design considering the properties of physical materials and the surrounding environment.  (b)~\systemName's spatial modeling interface. (c) Scrap-wood chair designed in our system.}
  \Description{(a) -- (b) --- (c)---}
  \label{fig:teaser}
\end{teaserfigure}

\maketitle

\section{Introduction}
Carpentry is an ancient craft that has evolved through centuries, integrating various technologies to enhance precision and creativity. Recently, digital fabrication has further expanded this field, enabling complex designs \new{\cite{zoran_human-computer_2013}}, efficient material usage \new{\cite{zhao_co-optimization_2022, wu_carpentry_2019}}, and automation \new{\cite{lipton_robot_2018}}. Despite these advancements, the needs of manual woodworkers, who either work with limited materials and/or want to design a custom assembly that fits into their environment, are not as widely addressed in human--computer interaction (HCI) research. Although several works offer alternatives to handcraft \cite{larsson_tsugite_2020, tian_matchsticks_2018, saul_sketchchair_2010, baudisch_kyub_2019, sethapakdi_fabricaide_2021, bourgault_coilcam_2023}, we aim to contribute to the body of work that complements existing manual workflows \cite{leen_jigfab_2019, toka_adaptable_2023, schoop_drill_2016, tian_adroid_2021}. For woodworkers, a primary challenge is planning a design that can balance competing considerations such as available material, space, tools and abilities. \textit{Do it yourself} (DIY) woodworkers, who we define as non-professionals with limited access to materials and/or tools, experience this challenge extensively because of added fabrication constraints~\cite{bezci_it_2016}.

Therefore, an effective computational support system for manual woodworking should encode wood-specific design methods and material considerations. Nordmoen et al.'s observations in wood workshops highlight a material-first approach \cite{nordmoen_making_2022}. For example, one shop took extra care to conserve timber, utilizing material that would otherwise be considered flawed for furniture creation and discarded. Physical interactions with the material are important, as they set woodworking apart as a \textit{craft}, rather than a purely industrial practice \cite{bunn_importance_1999, hoadley_understanding_2000}. To avoid using expensive material or inaccessible tools, DIY woodworkers must adopt unique design principles such as employing basic craft techniques and utilizing on-hand materials \cite{camburn_principles_2018}. More specifically, DIYers frequently use discarded wood (scrap) \cite{bonvoisin_design_2017}. 
They may already have access to a collection of wood off-cuts from past projects, or may gather scraps (often for free) from construction yards, maker-spaces, or online listings. 

Although acquiring this material is easy, planning a design around it is challenging. Scraps are random in terms of their size, appearance, and quality. \new{This requires woodworkers to adopt a \textit{bricolage} approach -- adapting their designs to fit materials on hand, as explored in previous HCI fabrication research \cite{efrat_hybrid_2016, beltagui_bricolage_2021, stemasov_ephemeral_2022}}. When successful, this approach is not only affordable, but aligns with \textbf{sustainable} practices, recycling material that would otherwise be wasted \cite{besserer_cascading_2021}. Several previous works focus on various types of scrap materials, including leather \cite{lin_leatherboard_2024}, plastic bags \cite{deshpande_upcycling_2022}, e-waste \cite{lu_ecoeda_2023}, 3D-printed filament \cite{wall_scrappy_2021, dew_designing_2019}, and \new{household waste \cite{wall_substiports_2023}}. We aim to contribute to this field with a creative design interface for discarded wood. Screen-based desktop computer-aided design (CAD) interfaces are not optimized for material-informed design because they are abstracted from the real world \cite{menges_material_2012}. We see an opportunity for an extended reality (XR) interface to connect physical materials to a virtual design in the same space, bringing \textbf{material awareness} into digital design. XR is also suitable for \textbf{in situ design}, allowing the woodworker to verify that a design is functionally and aesthetically compatible with their surroundings.

Therefore, we present \systemName, a system designed to support \new{functional} woodworking with discarded material \new{(\cref{fig:teaser})}. \new{\systemName is designed for makers who would rather \textit{adapt} their project to available scraps rather than purchase new materials for a fixed design.} Our system leverages the capabilities of a commercially available headset to create an XR experience that integrates into manual workflows. \systemName's backbone is a digital inventory of scrap wood containing a virtual \textit{twin} of each available scrap. Users interactively cut and join the pieces from this inventory into a virtual assembly. Specialized modeling utilities facilitate physically valid cuts, part placement, simple connections and snapping to the surrounding environment. While the user designs their assembly in XR, the system generates a 2D \textit{cut plan} for each twin. The system also provides real time feedback on material availability, such as informing the user when they attempt to extend the length of a cut piece beyond the dimensions of their material. This enables \textbf{creative negotiation} between material and design informed by environmental fixtures. Once the user finishes their design, they can view it on a screen-based interface with necessary information for fabrication. We demonstrate the feasibility of our system in collaboration with a DIY woodworker, where we design and build furniture using random scraps. We also share feedback from eight hobbyist woodworkers to determine how \systemName could integrate into their design process. %We show our results and share insights on how the system operates in practice. 

\vspace{0.2cm}

Our contributions include:
\vspace{-0.1cm}
\begin{itemize}[leftmargin=0.5cm]
  \item \new{A study of the requirements for DIY woodworking through formative interviews with six DIY woodworkers.}
  \item \new{An XR system for in situ visualization and design of functional assemblies built with on-hand materials, including a virtual inventory of 3D scraps, woodwork-specific modeling utilities, and user-controlled cut plan generation.}
  
  \item \finaledit{A case study and feedback sessions with target users demonstrating how \systemName supports the improvisational nature of woodworking with scraps.}
  % \item \new{A case study and feedback from target users highlighting how woodworkers might engage with the \systemName workflow in practice.}
  %\item We describe the system architecture and interaction design used to support users in creating a spatial design and cut pattern informed by physical properties of available wood. 
\end{itemize}

% enabling more creative negotiation between available materials and design. 

% \vspace{0.2cm}
% In summary, we make the following contributions:
% \vspace{-0.1cm}
% \begin{itemize}[leftmargin=0.5cm]
%     \item contribution 1
%     \item contribution 2
%     \item contribution 3
% \end{itemize}

\section{Related Work}

% This section discusses existing interactive design tools for woodwork (\cref{sec:related_work:interactive_design_for_woodwork}), ways researchers have integrated material constraints into design systems (\cref{sec:related_works:designing_with_material_constraints}), interfaces for situated design (\cref{sec:related_works:designin_in-situ}), and the use of XR for physical material interactions (\cref{sec:related_works:XR_for_physical_fabrication_and_design}).

\subsection{Interactive Design for Woodwork} \label{sec:related_work:interactive_design_for_woodwork}
There are several computational approaches that either augmented or replaced traditional CAD tools to facilitate the design of wood structures, with scopes ranging from joint-specific design ~\cite{larsson_tsugite_2020, yao_interactive_2017, magrisso_digital_2018} to general wood assemblies ~\cite{lau_converting_2011, schulz_design_2014, shao_dynamic_2016}. For example, \textit{Carpentry Compiler} provides real-time feedback on manufacturability for wood furniture designed with CAD, optimizing the plan after the design is finalized \cite{wu_carpentry_2019}.  \textit{MatchSticks} supports improvisational fabrication with a CNC machine~\cite{tian_matchsticks_2018}. These design interfaces focus on fabrication constraints. Other studies reported woodworking interfaces incorporating  validity~\cite{saul_sketchchair_2010, umetani_guided_2012} and reproducible workflows~\cite{tran_oleary_tandem_2024}. We share a common goal with these works in making digital design \textbf{\textit{less abstract}} through constraints; however, our work specifically addresses design with limited non-standard materials.
Further, some design interfaces are aimed at hobbyists who want to avoid complex CAD interfaces, who are also are target users. For example, \textit{CraftyAmigo}\footnote{\textit{CraftyAmigo} \url{https://www.craftyamigo.com}} allows users to snap together pre-defined wood parts to design furniture and provides a shopping list of necessary stock material. \textit{SketchUp}\footnote{\textit{SketchUp} \url{https://www.sketchup.com/en/products/sketchup-for-web}} is another easy-to-use tool that allows a user to generate a cut list after creating a virtual model. However, the pattern generation is separated from the design process and is assumed to work with perfect stock. As desktop applications, these tools are inherently disjointed from physical material and surroundings. In contrast, our system enables \textit{situated design}, allowing users to make material-informed design choices in situ.

\subsection{Designing with Material Constraints} \label{sec:related_works:designing_with_material_constraints}
In CAD workflows, a material is typically treated as “a passive property of form, rather than as an active form-generator,” which means that the material is assigned to the completed design instead of being embedded into the design process \cite{menges_material_2012}. There is a growing body of HCI research that addresses this gap between material and digital design. \new{Previous material-first design methods have targeted a variety of materials such as clay \cite{bourgault_coilcam_2023}, stone \cite{johns_framework_2023}, and organic tree branches \cite{mollica_zachary_tree_2016}.} Material constraints may also augment design for rectilinear material, such as laser-cut assemblies \cite{baudisch_kyub_2019, dogan_sensicut_2021}. \textit{Fabricaide}, a pattern generation system for laser-cutting, shares our goal of providing real-time feedback on material availability while the user designs their project \cite{sethapakdi_fabricaide_2021}. For manual woodworking, some works packed parts onto user-defined stock by optimizing for material cost and time \cite{leen_jigfab_2019, wu_carpentry_2019, zhao_co-optimization_2022}. Although auto-packing approaches are efficient, they necessitate perfect materials and limit opportunities for creative design \textit{around} material properties.  Koo et al. took a unique approach by suggesting modifications to the 3D furniture design for minimizing waste as they auto-packed the pattern \cite{koo_towards_2017}. Instead of algorithmic optimization, we generate patterns \textbf{interactively} similarly to \textit{PacCAM}, which uses 2D rigid bodies and snapping for easy arrangement~\cite{saakes_paccam_2013}. This user-controlled packing approach material informed design decisions beyond surface conservation. %\textit{PacCAM} registered material textures using a camera, while we use procedural textures to emphasize physical material interaction. 
Since we focus on rectilinear wood, we also forgo the need to create 3D scans, which have been utilized in other works~\cite{cousin_wild_2023, larsson_human---loop_2019, sunshine_inventory_2022}. 

\subsection{Designing In Situ} \label{sec:related_works:designin_in-situ}
In situ interfaces introduce spatial awareness and context-specific data into the design process, which are otherwise lost in screen-based software. Existing systems have brought situated design to a variety of fields, including plastering \cite{mitterberger_interactive_2022}, circuits \cite{kim_virtualcomponent_2019} and ergonomic desks \cite{lee_interactive_2018}. To assist furniture design, \textit{Protopiper} enables users to physically prototype large-scale objects through plastic tube extrusion \cite{agrawal_protopiper_2015}. Lau et al. provided a method to design “complementary” household objects using a photo reference \cite{lau_modeling--context_2010}. Other works turned existing objects into design tools, such as physical “stamps” \cite{lau_situated_2012}, traced shapes \cite{reipschlager_designar_2019}, or scanned meshes \cite{weichel_mixfab_2014}. Similarly, we provide design guidance using to-scale objects; however, instead of physical material, users assemble with virtual “twins.” Stemasov et al. utilized XR to preview design alterations to existing artifacts \cite{stemasov_mixmatch_2020, stemasov_brickstart_2023}. Their work, \textit{pARam}, provides a real-time AR preview of adjustments made to existing parametric design models \cite{stemasov_param_2024}. Although easy to use, suitable parametric models can be difficult to source, so we enable users to create custom designs directly in XR via woodworking-specific modeling utilities. Finally, we acknowledge that real world measurements are crucial for creating designs that fit in space and accommodate available material \cite{ramakers_measurement_2023}. \textit{SPATA}'s physical tools measure real-world objects and encode the data into CAD software \cite{weichel_spata_2015}. Instead of specialized hardware, we use existing dimensions of uncut material and spatial data obtained using our XR headset to determine the final measurements for assembly.

% \finaledit{\textit{Draw2Cut} allows users to fabricate artifacts without measurements by sketching cuts directly on physical materials using a custom hardware 
% setup~\cite{gui_draw2cut_2025}.} 

\subsection{XR for Physical Fabrication and Design} \label{sec:related_works:XR_for_physical_fabrication_and_design}

% Recent research demonstrates how XR systems can interact with physical objects in substantial ways, including in situ spatial search \cite{stemasov_shapefindar_2022}, sketched physical interaction \cite{kaimoto_sketched_2022, suzuki_realitysketch_2020, xia_realitycanvas_2023}, ambient overlays \cite{gonsher_telepresence_2024, han_blendmr_2023}, context-aware layouts \cite{cheng_semanticadapt_2021, lindlbauer_context-aware_2019}, and robotic intervention \cite{suzuki_roomshift_2020, ihara_holobots_2023}.

There is a growing body of fabrication-focused XR systems that augment interactions with real objects and/or materials\finaledit{~\cite{dogan_fabricate_2022}}. \textit{ToolDevice}, one of the earliest examples of XR handcrafting, enabled users to build virtual wood models using physical tools \cite{arisandi_virtual_2014}. \textit{Knock on Wood} also explored tool-based virtual interactions in XR \cite{strandholt_knock_2020}. Yanpanyon et al. demonstrated that XR improves spatial reasoning for DIY furniture assembly \cite{yanpanyanon_spatial_2020}. The architectural startup \textit{Fologram}\footnote{\textit{Fologram} \url{https://fologram.com/}} provided an XR framework for architectural-scale fabrication. Cousin et al. used \textit{Fologram} to construct structures with nonstandard wood branches \cite{cousin_wild_2023}, while Jahn et. al used the system to fabricate structures with reclaimed timber offcuts \cite{jahn_mixed_2024}. Further, Parry et. al utilized \textit{Fologram} to reuse scrap timber, using a holographic overlay to cut nested parts and assemble them into a structure \cite{parry_recycling_2021}. These \textit{Fologram}-based works demonstrate that XR is capable of supporting physical assembly. They focused on XR as a fabrication guide, and designs were generated with computational models in advance. \new{Most adjacent to our work, Kyaw et al. utilized \textit{Fologram} for parametric exploration of bamboo's bending properties in situ \cite{kyaw_active_2023}.} In contrast, our system \new{supports creative scrap wood design for personal assemblies}.

\section{Formative Interviews}

In this section, we detail our findings from a series of formative interviews focused on the needs and challenges of DIY woodworkers. Due to their lack of professional training, we assume they are more likely to underestimate material requirements and not understand how their design will function in a given space. We also assume that DIYers are more likely to use scrap material and basic tools. We interviewed six non-professional woodworkers (P1-P6), between ages 22 to 71, to evaluate these assumptions and enrich our practical understanding of how they execute projects. \new{Participants were unpaid volunteers recruited from the local maker community. P4, P5, and P6 prefer to use scrap material for their DIY projects, while the other participants use scraps on a per-project basis. P3 and P5 rely on CAD to create woodworking plans, while the rest of the participants work with paper.} We conducted semi-structured interviews that lasted approximately 1 hour. The first 45 minutes were spent on design approaches, tools and material. The last 15 minutes were dedicated to discussing spatial computing use cases for woodworking, and \new{those unfamiliar with XR  were shown examples of 3D objects in a mixed-reality environment.} \new{We took a bottom-up approach in our thematic analysis of interview responses to identify three main challenges:} managing material needs (\cref{sec:formative_interviews_material_needs}), negotiating between material and design (\cref{sec:formative_interviews_negotiating}), and spatial visualization (\cref{sec:formative_study:in-situ}).

 % Only P1 and P2 had experience using an XR headset.
\subsection{Managing Material Needs} 
\label{sec:formative_interviews_material_needs}
All interviewees reported struggling with either the high cost of quality timber or limitations of available scrap materials. P1, who was the president of her university's makerspace, noted a reliance on scrap wood due to financial constraints. The makerspace treats scrap wood as a shared resource, typically adequate for low-cost, functional designs. P1 also observed that students working with scraps often use hand and power tools, reserving digital fabrication tools for more precise and expensive projects. P5 mentioned that it takes time to determine the suitability of discarded material, requiring him to touch and closely examine every scrap to \textit{"fully understand"} it. Five out of six interviewees  reported that they have had to manually adjust a design after starting to cut and assemble it because they inaccurately predicted their material requirements. P2 stated, \textit{“I always underestimate how much lumber I need... [wood] is not infinite.”} P4 explained that he always has \textit{“90\% of the material [physically with him] straight away”} before designing his assembly to be certain about not falling short.

\subsection{Negotiating Between Material and Design} \label{sec:formative_interviews_negotiating}
We observe two approaches to mapping a design to available materials from our interviews: 1) a material-first approach that involves creating a design to accommodate material that has been allocated for a project, and 2) a design-first approach that involves searching for material to accommodate an existing plan. In practice, these approaches often overlap. For example, P1 was building a bathroom shelf and initially planned for three shelves. After realizing she had more material than expected, she added two extra shelves. Such improvisational decisions, while successful in P1's case, are not always feasible mid-assembly. 
To create a \textit{cut list}, which details the dimensions and quantities of pieces needed for a project, P2 sketches out necessary pieces and mentally \textit{“extrapolates from the sketch,”} while P5 manually inputs measurements into CAD and creates a digital model, before disassembling it into a 2D pattern, printing it out, and tracing cuts onto available wood. 

Although there are online cut-list optimizers, P6 explained that these tools are incompatible with his typical workflow because it is \textit{“rare that [he] has a lot of stock material,”} on hand, and that it he is also \textit{“unsure if [he] can specify the parts he needs in advance [of the design]”}. As woodworkers organize their cut lists, they often dynamically edit their design; P6 likens this process to “\textit{sketching}.” Conserving surface area is not the only goal throughout this process: the physical logistics of fabricating cuts must also be considered. P1 creates patterns to avoid repeatedly re-calibrating her table saw, while P3 and P6 determine how precise their patterns are based on the tools at their disposal and their self-perceived skill levels. Further, they juggle other material properties as they design: P1 said she would appreciate a tool that allowed her to design around imperfections, thereby creating less waste. 

% \vspace{0.2cm} \noindent
\subsection{Spatial Visualization} \label{sec:formative_study:in-situ}
Spatially visualizing an assembly while it is being designed is a bottleneck voiced by all interviewees. P5 relies on design software but picks up wood scraps and \textit{“physically rotates them in mid air”} to understand how they occupy space before switching to designing on his computer. All interviewees acknowledged that spatial computing could enhance this process. P2, for instance, designed planters on paper but realized post-assembly that they \textit{“visually dominated”} his backyard. He suggested that if he owned an XR headset, he could have used it to visualize and adjust his plans. P4, a cafe owner who built a new counter-top with reclaimed wood, found that the final layout made accessing his fridge difficult. Previewing his design in situ would have enabled him to practice moving around different configurations and adjusting the design accordingly, instead of realizing its limitations post-assembly. 
\subsection{Takeaways}
\new{DIY woodworkers must estimate and manage material requirements over the entire life cycle of a project. Even when all materials are already available on-hand, predicting if it is sufficient for their intended project remains difficult. They must also determine if available material meets structural and aesthetic standards. This is complicated by the need to balance multiple material considerations at once, especially when irregular scraps are involved. Translating a design to physical materials requires mental extrapolation, and going back and forth between design creation and material allocation requires intensive context-switching. Further, we observe that it is difficult to explore the material-constrained design space without making irreversible cuts. Woodworkers find it difficult to visualize how a design will occupy space within its environment, whether it be in terms of physical dimensions, layout, or functionality. Woodworkers who use CAD (as opposed to sketching) could achieve a more precise spatial understanding of their design; however, re-contextualizing a digital model for the physical material and space is tedious. In general, a lack of situated spatial visualization leads to post-assembly adjustments, wasted material, or dissatisfaction with the built object.}
Finally, P1 also shared that her university's maker-space just opened a new XR division where students can borrow headsets. This is in line with our vision of XR becoming a practical and accessible addition to DIY craft.

\section{\systemName}
% \section{\systemName: Material-Aware In Situ Design of ...}
% \systemName introduces an advanced mixed reality platform designed specifically for the woodworking domain. This system merges real-time spatial computing with an intuitive user interface, enhancing the traditional woodworking process by enabling precision and creativity in design and execution.

% \new{In this section, we provide an overview of \systemName (\cref{section:system:overview}) and introduce system components: registration of physical materials (\cref{section:system:physical-material-registration}), spatial modeling utilities (\cref{section:system:spatial-modeling}), linking the cut-plan and 3D assembly (\cref{section:system:linking}), and the screen-based interface for fabrication (\cref{section:system:design-preview}).}

\subsection{Overview} \label{section:system:overview}
\begin{figure}[h]
  \centering
  \includegraphics[width=.99\textwidth]{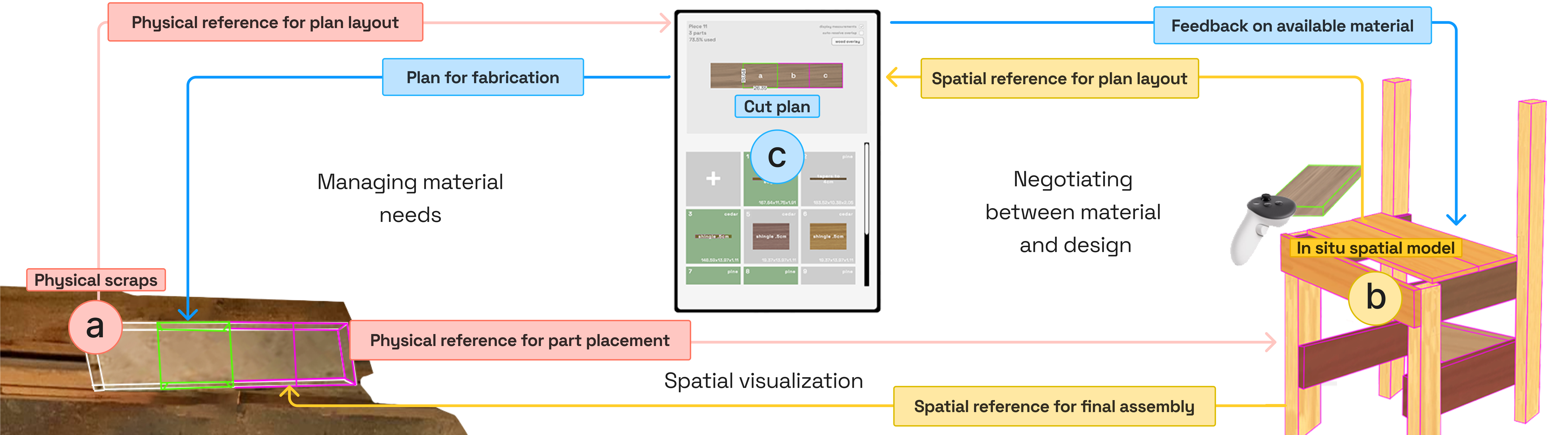}
  \caption{\systemName's three woodworking design contexts. We label each context interaction with the challenge it addresses, as identified from our formative interviews.}
  \Description{(a) -- (b) --- (c)---}
  \label{fig:contexts}
\end{figure}

Based on the challenges identified by our formative interviews, we propose an XR system that connects three woodworking design contexts: 1) the \textbf{physical material}  (\cref{fig:contexts}a), 2) in situ \textbf{spatial model}  (\cref{fig:contexts}b), and 3) \textbf{cut plan} (\cref{fig:contexts}c). Together, these contexts help a woodworker engage in an iterative feedback loop towards a design informed by real constraints. Unlike traditional workflows, \systemName tracks all three contexts simultaneously, supporting on-the-fly design decisions as new material constraints and spatial considerations emerge. 

\new{The \systemName workflow is divided into three phases: \textbf{material registration}, \textbf{design}, and \textbf{fabrication}. The design phase, which involves spatial modeling and cut-plan generation, occurs in XR. Material registration and fabrication occur out of the headset, aided by our screen-based interface. The system supports back-and-forth transitions between the material registration and design phases; however it assumes that design \textit{intent} is finalized before physical fabrication.}

\subsection{Material Registration} \label{section:system:physical-material-registration}

\begin{figure}[t]
  \centering
  \includegraphics[width=.75\textwidth]{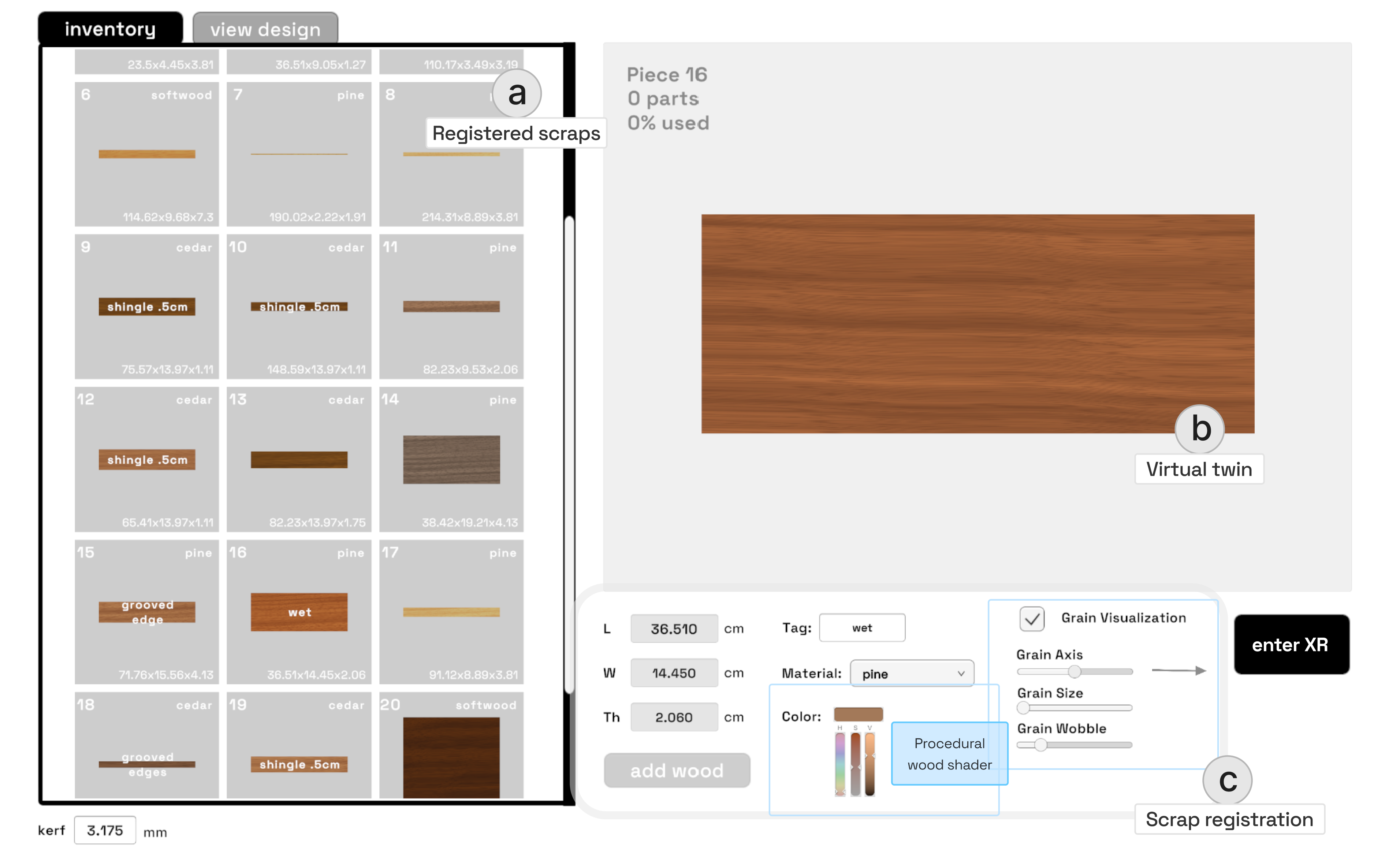}
  \caption{Virtual inventory and scrap registration interface.}
  \Description{(a) -- (b) --- (c)---}
  \label{fig:registration}
\end{figure}

%Before a user begins to design a project, they need 
The first step is to register available materials into \systemName's \textbf{virtual inventory} using a screen-based interface such as a desktop or laptop computer (\cref{fig:registration}). 
\new{The inventory may persist across design sessions, and scraps can be re-registered on the screen-based interface during any point of the design process; however, we assume that the user will start with some registered material.}
%This step takes place on desktop computer, allowing the user to take measurements and type values outside of the headset.
To register a scrap, the user manually measures its dimensions and types the values into our interface. They may add an optional tag to note intended usage, imperfections, or other relevant features. Further, the user can assign a material type, associated with a predefined set of parameters driving the system's procedural wood shader. The user may choose to edit grain axis, size, wobble and color to better match the scrap. %We created a fully procedural wood shader to capture the natural range of appearance in diverse wood types. 
%\cref{fig:registration}b shows the four parameters that can be controlled by the user on a continuous scale. 
Once a scrap is registered, its thumbnail appears in the inventory, along with its tag, material type, and dimensions. Each virtual scrap is assigned a number, and the user will mark the same number on the physical scrap to link the two.

\subsection{\textbf{Spatial Model}} \label{section:system:spatial-modeling}

\begin{figure}[t]
  \centering
  \includegraphics[width=.95\textwidth]{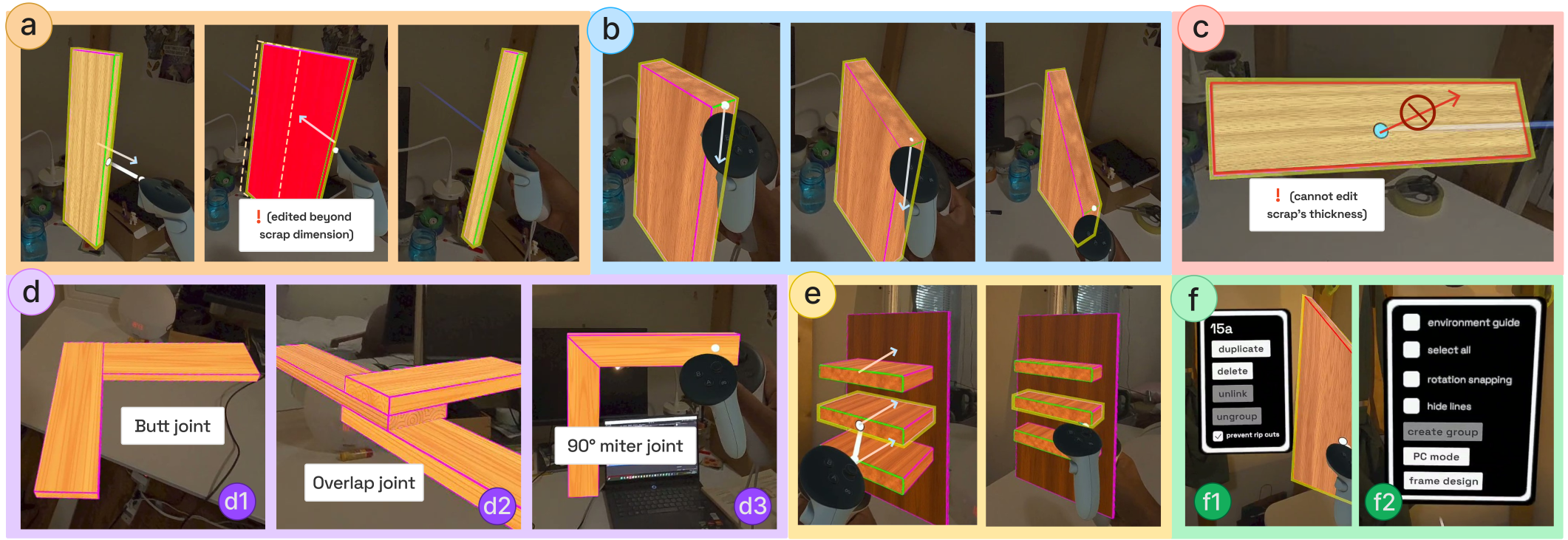}
  \caption{Spatial modeling utilities provided by our system. (a) Face manipulation. (b) Edge manipulation. (c) Restricted resaw cut. (d) Bare cut joints made with our system. (e) Linked cuts being edited simultaneously. (f1) Part options. (f2) Scene options.}
  \Description{(a) -- (b) --- (c)---}
  \label{fig:spatial}
\end{figure}

Each inventory entry references a virtual \textbf{twin} of the physical scrap, and become accessible once the user wears their headset. The user assembles these twins into a spatial model with the assistance of woodworking-relevant modeling utilities. \new{Certain utilities described below are accessed via canvas menus (\cref{fig:spatial}f)}.

\paragraph{\textbf{Cutting Parts}}
To add a scrap to their design environment, the user opens the virtual inventory menu in XR and selects its thumbnail, causing a 3D to-scale virtual twin to appear in front of them. The user can then subtract material from the twin by directly manipulating faces and edges, thereby cutting a \textbf{part}. Faces are selected with a \textit{ray interactor} (a pointer that extends out from a controller to a selection surface), and can be pushed or pulled along their normal vector (\cref{fig:spatial}a). Edges are selected with a \textit{poke interactor} (a point on the controller that directly touches a selection surface), and can be moved along one of two perpendicular axes (\cref{fig:spatial}b). If the user attempts to return an edge back to $90^\circ$, it will snap into place within a small tolerance of being square. 

The user either duplicates an existing part or selects the twin's 2D cut plan to cut new parts out of a twin. These parts can be resized, duplicated, and deleted either from the cut plan, or by selecting the part in 3D space. By default, our system restricts resaw cuts (\cref{fig:spatial}c), which split a piece of wood along its thickness, because they are difficult to execute without advanced tools. Users can disable this constraint if desired. %Since cuts are made by moving edges and faces, all resulting parts are eight-cornered and rectilinear.

\paragraph{\textbf{Positioning Parts}}
Parts are moved through \textit{grab} interactions. The user grabs a part by squeezing the controller while touching the part, and the part will move and rotate with the controller until the user releases it. \new{Our system uses physics-based collisions to automatically detect interactions between parts and applies a force to resolve solid intersections in 3D}. %During development, we found that the controller's natural rotation is unconstrained and difficult to control.To address this, 
We also provide a utility that snaps the rotation of a part to $15^\circ$ increments across three dimensions. $15^\circ$ provides the range necessary for complex designs while staying easy to control.

\paragraph{\textbf{Joints and Linked Parts}}
Our interface supports bare cut joints (\cref{fig:spatial}d), including butt joints, overlap joints and mitered butt joints. For miters, we provide feedback when an edge is edited to form half of a joint of a user-specified angle. Our system also provides support for repeating parts through \textbf{linked} editing (\cref{fig:spatial}e). All parts in the same linked group experience the same edits in real time (though their positions are \textbf{not} linked). For example, a miter joint may only require editing one edge to create two compatible parts. Linked parts are indicated by shared highlighting across each part when an edge or face is selected.

\begin{figure}[h]
  \centering
  \includegraphics[width=.95\textwidth]{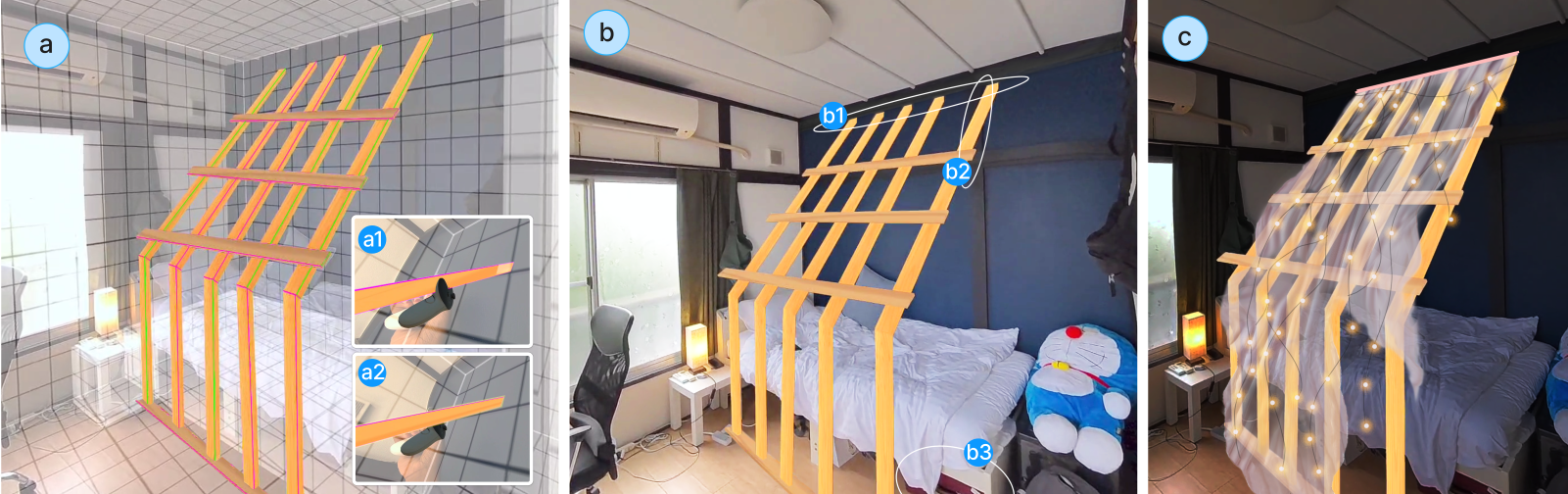}
  \caption{We used the scene mesh to design a bedroom canopy in situ. (a) The canopy design and scene mesh. (a1) A part intersects with the scene mesh. (a2) An intersection with the scene-mesh is resolved once the user releases the part. (b) The canopy design without the scene mesh. (b1) Top of the canopy frame intersects with an existing horizontal beam along the wall, and therefore, can be fastened to the beam when fabricated. (b2) The frame avoids an unwanted intersection with an existing vertical beam. (b3) The design of the frame provides enough space to push a piece of luggage underneath the bed. (c) A photo-illustration imagining the design in use.}
  \Description{(a) -- (b) --- (c)---}
  \label{fig:scenemesh}
\end{figure}

\paragraph{\textbf{Scene Mesh}}
Similar to part-to-part interactions, we use physics-based collisions to prevent overlap between parts and the \textbf{scene mesh}, \new{as seen in \textit{SnapToReality} \cite{nuernberger_snaptoreality_2016}}. A scene mesh is a spatial representation of the surfaces of the user's workspace, including walls, floors, tables, and other furniture. This mesh is created through in the scene-setup application of the headset. The scene mesh is automatically loaded into \systemName's spatial design interface, and it provides reference surfaces for part alignment, preventing an assembly from "floating" in space. \cref{fig:scenemesh}a shows the scene mesh bounding a canopy frame design.  

\subsection{Cut Plan}

\begin{figure}[h]
  \centering
  \includegraphics[width=.95\textwidth]{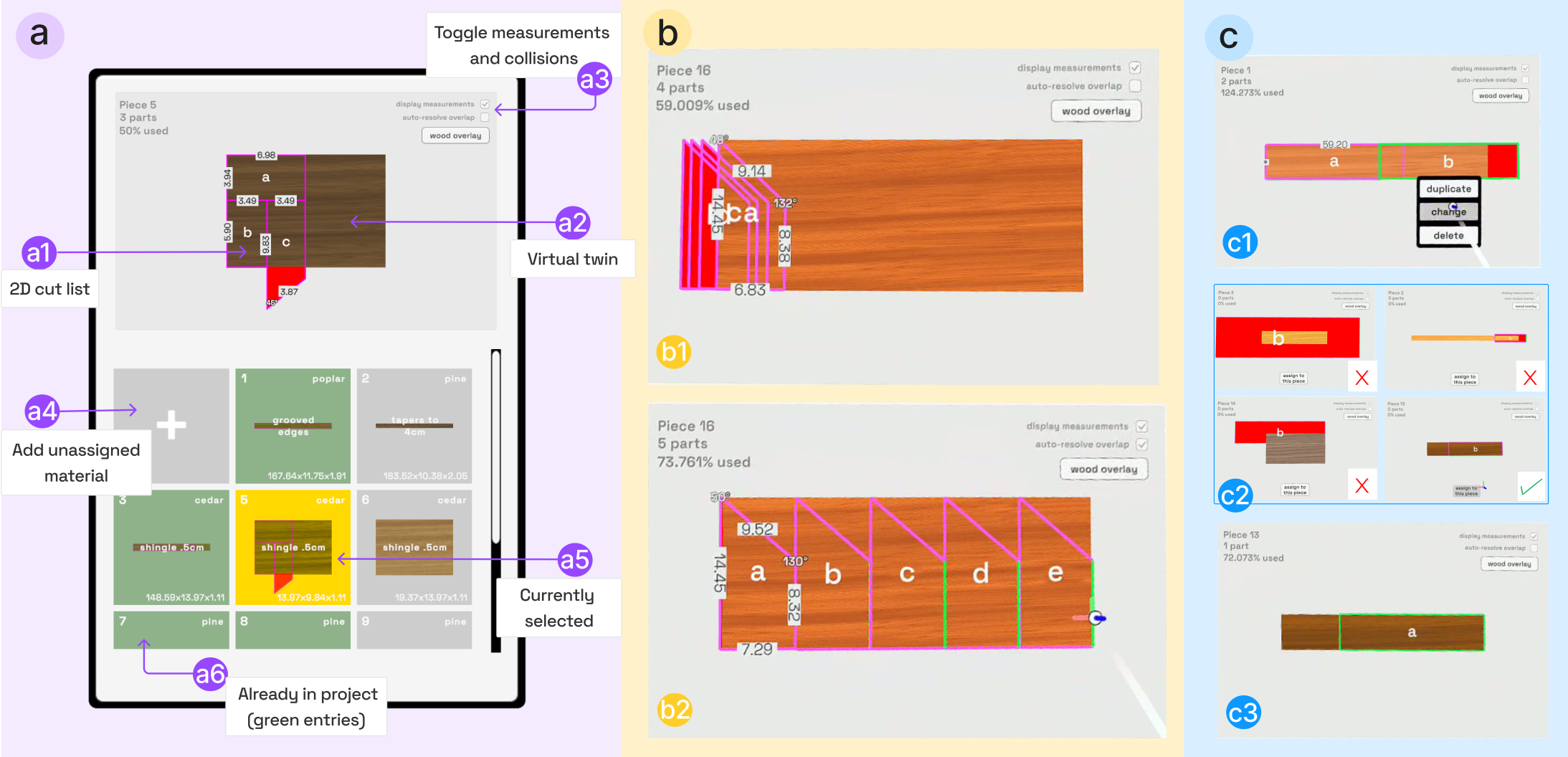}
  \caption{(a) Cut plan interface. (b) Our two collision modes: (b1) Parts overlap in the \textit{manual} collision mode. (b2) Parts redistribute themselves in the \textit{auto-resolve} mode. (c) Reassigning parts. (c1) Part b is too big for scrap \#1. (c2) Previewing if part b fits onto other scraps. (c3) Reassigning part b to scrap \#13.}
  \Description{(a) -- (b) --- (c)---}
  \label{fig:menu}
\end{figure}

As users cut twins into parts and assemble them into a design, \systemName generates an interactive \textbf{cut plan} (\cref{fig:menu}a). This plan is presented like a map showing the layout of parts on each scrap.
%, with each 3D part in the spatial model appearing as a 2D shape on the plan. 
%This part's 2D position on the twin indicates where the user will make their cuts. 
The plan is displayed in 2D as default because we expect that users will not typically make resaw cuts. When needed, they can rotate the cut-plan $90^\circ$ around its y-axis. 

\paragraph{\textbf{User-Guided Packing}}
%Instead of automatically optimizing the layout, we expect users to prefer having control over cut placement, as indicated by our formative interviews.
To support interactive packing, we use 2D collisions for automatically preventing overlap between parts. We add a small distance (e.g., 3 $mm$) between parts to account for \textbf{kerf}, which is the material removed by the width of the blade when cutting. Users can customize the kerf-value based on their tools.
Collider-aided packing can be utilized in either auto-resolve or manual mode. 

In \textbf{auto-resolve} mode, cuts automatically resolve their intersections in real time (\cref{fig:menu}b2). As users create new 3D parts, each corresponding 2D part will be pushed into free space on the pattern, only overlapping when space is unavailable. We expect users to start making cuts in the auto-resolve mode and then switch to the manual mode as they move towards completing their design. 

In \textbf{manual} mode, parts do not resolve their intersections until they are selected and moved around on the pattern (\cref{fig:menu}b1). This mode is ideal for the precise placement of individual parts when the rest of the pattern is finalized. Users edit individual part placements on the cut plan by selecting a part on the 2D canvas and moving it around or rotating it in $90^\circ$ increments. The part will automatically snap to the edges of the twin if close. \new{Direct pattern manipulation can be performed in XR (with a ray interactor) or with a screen-based interface and a mouse for better precision.}

\paragraph{\textbf{Reassigning Parts}}
During the design process, a user may cut a part out of one scrap, before realizing it makes more sense to take it from a different material. Therefore, the cut plan provides a \textbf{reassign} feature that enables previewing a part to-scale next to other twins, along with any existing cuts on that scrap, allowing the user to assign the part to the best fit (\cref{fig:menu}c). The user can also opt to create \textbf{unassigned} parts that are not associated with any twins to estimate material they do not have on hand. %Unassigned parts will show up as bright red in the spatial model.

\subsection{Linking the Cut Plan and 3D Assembly} \label{section:system:linking}

\begin{figure}[H]
  \centering
  \includegraphics[width=.85\textwidth]{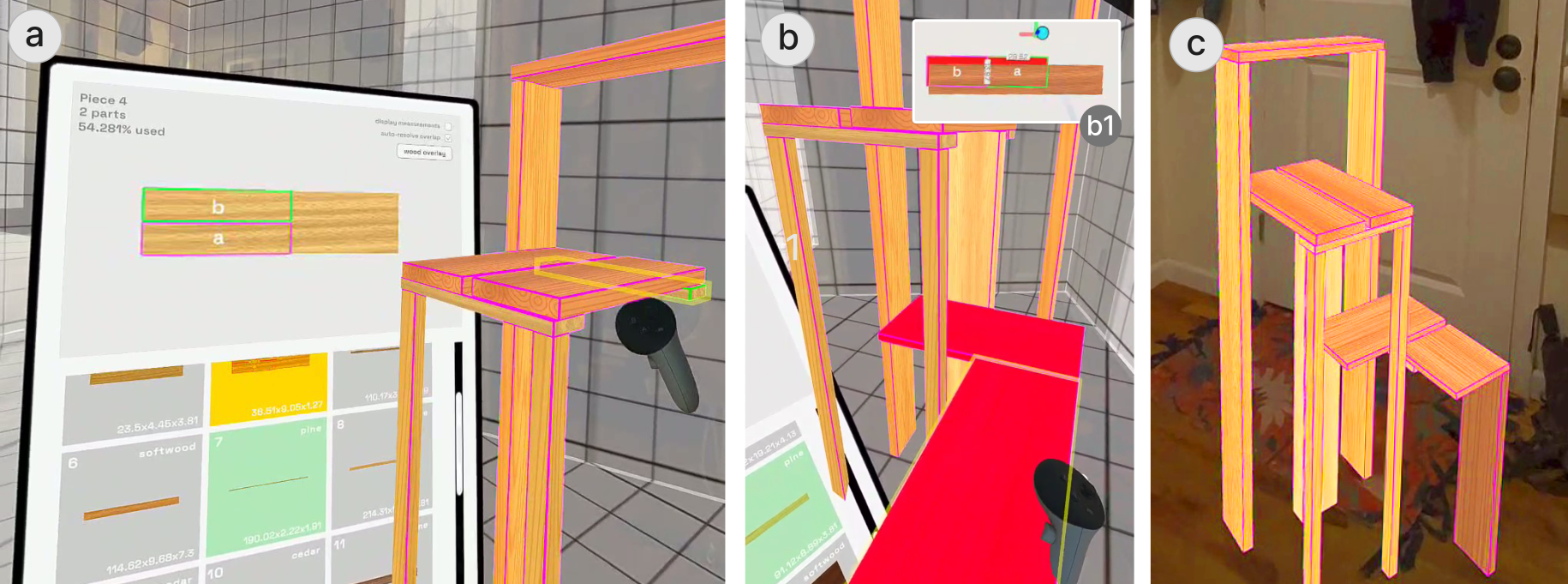}
  \caption{\systemName's material feedback helped us design a creative shelving unit. (a) A part selected on the spatial model is jointly highlighted on the cut-plan. (b) Warning coloring on invalid spatial parts, caused by (b1) cuts that exceed their material's boundaries. (c) The final unit designed in our system.}
  \label{fig:linked}
\end{figure}

\systemName provides real time, continuous feedback on material usage by linking the 2D cut plan with the 3D assembly. Each cut and its placement on the material are labeled with unique identifiers displayed in both 2D (on the cut plan) and 3D (within the spatial design interface). The system bridges these two views through a responsive highlight system (\cref{fig:linked}a).

\paragraph{\textbf{Feedback}} When a part is selected in XR, the corresponding scrap is selected in the virtual inventory, and the edges around the specific part are highlighted on the cut plan. Similarly, any adjustments made directly on the cut plan are reflected in 3D, with the same highlights. Before an adjustment is made, these highlights indicate what is about to change, helping the user verify that they are editing the correct edge or face.

The system provides feedback on material usage when new cuts are made and existing parts are adjusted. A warning system activates if a cut exceeds the available material boundaries or intersects with another part: the controller vibrates and the part turns bright red. The part will remain red, in both 2D and 3D, until the issue is corrected (\cref{fig:linked}b). Although most design edits are performed by manipulating parts in 3D space, certain situations necessitate selecting and resizing a part with a \textit{ray interactor} on the 2D cut plan. This is useful when working with parts that are difficult to access in space, such as a high shelf.

%Even if there is no direct violation, users may want to intermittently reference the 2D pattern to check if the proposed cut-plan is feasible. 

 %By selecting and resizing a part directly on the 2D pattern, the user can  adjust dimensions in relation to the overall cut arrangement.

\begin{figure}[h]
  \centering
  \includegraphics[width=.75\textwidth]{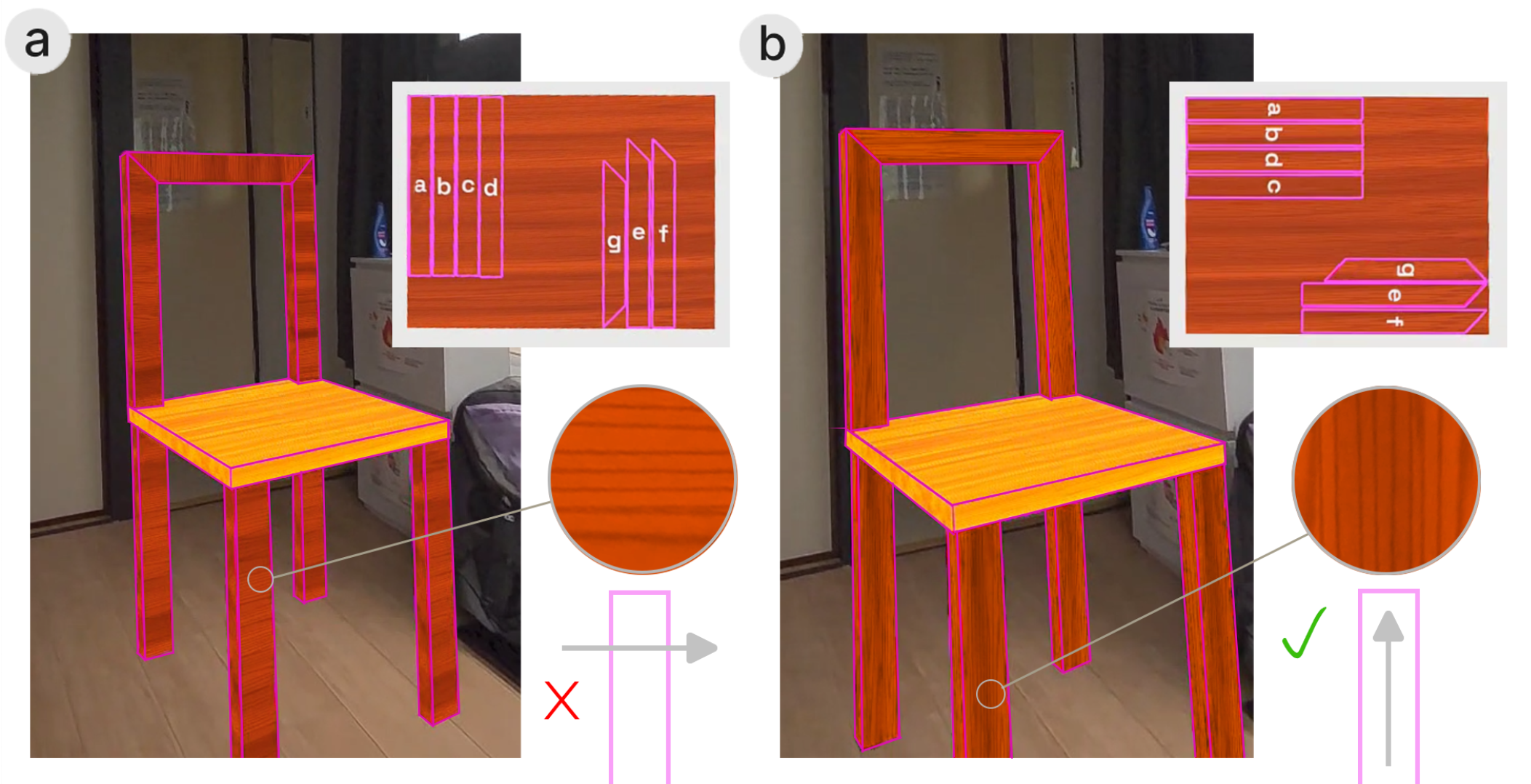}
  \caption{(a) Grain is perpendicular to the lengths of the chair's rails and legs, which is more likely to break. (b) Correct grain orientation after the pattern is fixed.}
  \Description{(a) -- (b) --- (c)---}
  \label{fig:grain}
\end{figure}

\paragraph{\textbf{Adjusting for the Grain Direction}} In addition to handling dimensions, the system enables users to adjust cuts based on the grain direction. The correct grain alignment is critical for structural integrity and aesthetic outcomes. The grain texture on each 3D part matches its relative 2D position on the scrap twin. If the grain direction of a part runs perpendicular to its length, the user can easily spot this and rotate the 2D cut accordingly (\cref{fig:grain}).

% \subsection{Aligned References} \label{section:system:physical-wood}

% By virtue of being in XR, our system allows users to engage with real wood during their design process. We expect users to naturally reference both the virtual inventory and their \textbf{physical inventory} when selecting suitable scraps for their design. %This is in line with the observation that tactile feedback is essential for raw material crafts \cite{crevels_coarse_2023}. 
% All design operations outlined above can be performed with a single controller, so the other hand is free to interface with physical material. We previously considered enabling hand-tracking, but it proved too imprecise for spatial assemblage. 

\begin{figure}[h]
  \centering
  \includegraphics[width=.95\textwidth]{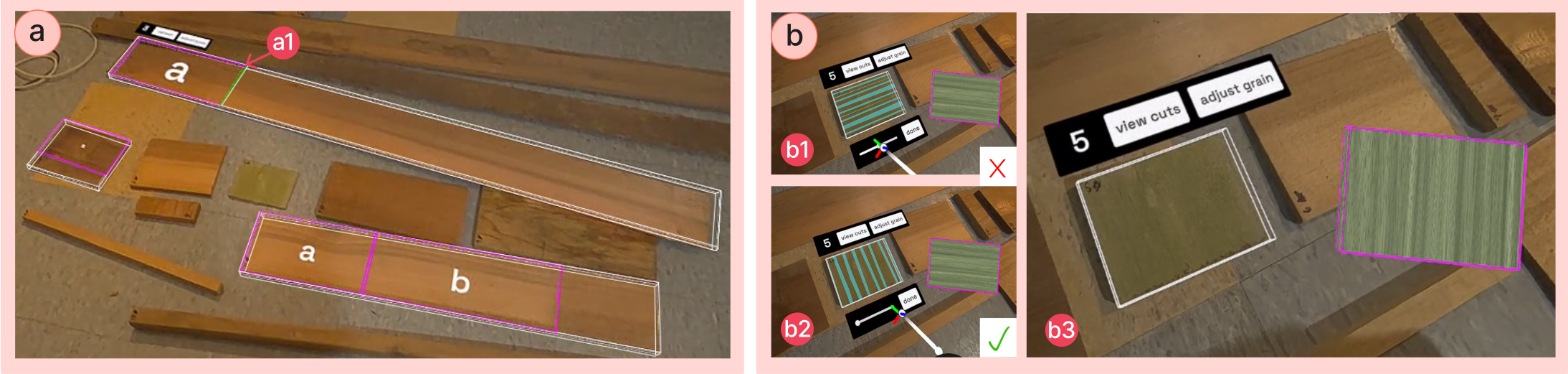}
  \caption{(a) Aligned references show proposed cut plans projected onto each scrap. (a1) An edge highlighted on the 3D assembly will show up on its part's reference. (b) Fixing the grain axis of a twin whose grain does not match its scrap. (b1) Original, incorrect grain axis. (b2) Corrected grain axis. (b3) Physical wood next to its corrected twin.}
  \Description{(a) -- (b) --- (c)---}
  \label{fig:ref}
\end{figure}

\paragraph{\textbf{Aligned Spatial References}} The user may opt to manually align a \textbf{virtual reference} to each physical scrap. The reference template can be aligned to a scrap laying flat on the ground with the assistance of the floor plane of the scene mesh. Once the reference is flat, the user may nudge the physical scrap into place. Once aligned, the user can select the physical scrap with a \textit{ray interactor} to either preview all proposed cuts on its surface (\cref{fig:ref}a) or re-draw the grain axis if incorrectly initialized (\cref{fig:ref}b). The cut preview can help users verify that planned cuts accommodate any physical surface features.

\subsection{Previewing the Design for Assembly} \label{section:system:design-preview}
\begin{figure}[t]
  \centering
  \includegraphics[width=.75\textwidth]{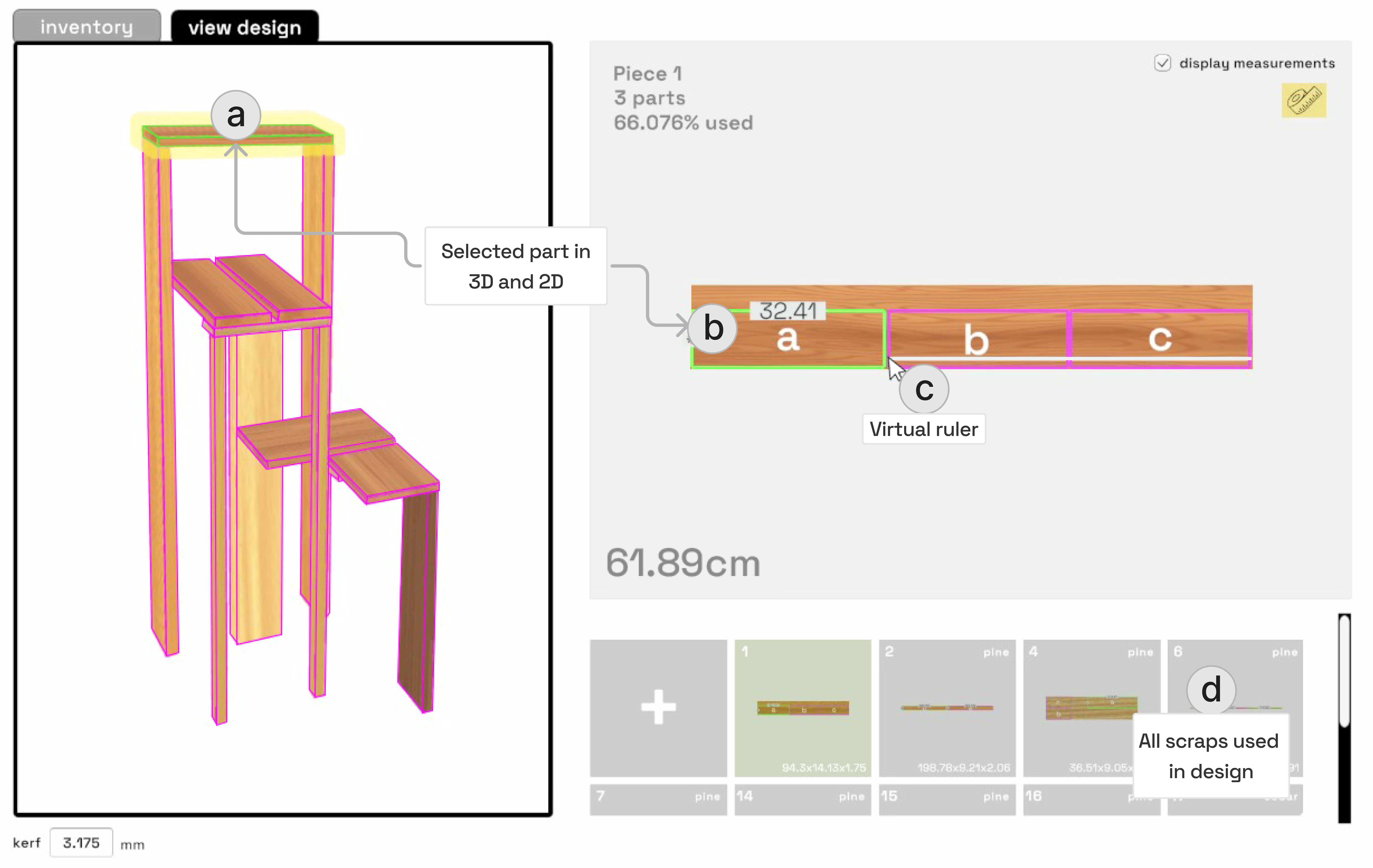}
  \caption{Design preview interface for fabrication.}
  \Description{(a) -- (b) --- (c)---}
  \label{fig:design}
\end{figure}

We provide a laptop/desktop interface that displays the entire 3D assembly and cut plan for users to fabricate their completed design (\cref{fig:design}). When a part is selected in the 3D assembly, its corresponding cut plan entry is highlighted, and vice versa. Users may need to utilize a virtual measuring tape to verify dimensions when parts are not aligned with their scraps' edges. \new{In the current implementation, a user cannot edit the spatial model from the screen-based design preview, but they can return to XR mid-assembly to revise their design in response to fabrication errors. They may also need to adjust their XR model after re-registering materials due to incorrect measurements or changes from planing or sanding.}

\section{Technical Implementation} 
\systemName is implemented in \textit{Unity 2022.3.31}, scripted in C\#, and is designed to operate on the \textit{Meta Quest 3}. We chose the \textit{Quest 3} for its widespread availability and full-color \textit{Passthrough} camera, which enables users to view their surroundings through the headset. Our XR utilities were developed using the \textit{Meta XR All-in-One SDK}. Geometry is generated at runtime using the \textit{ProBuilder Scripting API}. Interactive menus were built with Unity's \textit{UI Toolkit}, and custom materials were created with the \textit{Shader Graph}. We deploy \systemName as a \textit{PCVR} application to support the dual headset-PC interface. The headset must be connected to a computer to run the software, either through a link cable for reduced latency or via the \textit{Quest's} \textit{air-link} feature for a tether-free experience, depending on the user's preference.

\paragraph{\textbf{Collisions}}
Our system uses Unity's physics-based collision system for positioning 3D parts in spatial modeling and placing 2D cuts on the cut plan. We changed the default \textit{contact offset} that controls the precision of 3D collisions, to $10^{-7}$, which was the minimum value required for parts to stick together without gaps. 
%This ensures precise measurements when parts are resized to fit between other parts and/or the scene mesh. 
If a user grabs a 3D part and moves it into existing geometry, the system will temporarily allow the violation until the grab is released, thereby pushing the part out along the saved normal direction of the collision to resolve the intersection. On the 2D cut plan, every time a part is generated, resized, or re-positioned, we check for any existing overlap every frame, activating our warning systems accordingly. These 2D colliders assist users in resolving this overlap using Unity \textit{2D Rigidbodies}. We add kerf padding to the cut plan by dilating the colliders by half the blade width.

\paragraph{\textbf{Spatial Anchors}}
In situ design creation is supported by \textit{Meta's Mixed Reality Utility Kit's (MRUK)} spatial anchors, locking 3D parts into their user-defined positions in reference to the real world. If a user takes off their headset and walks away, their virtual assembly will remain in the same spot on their return. These anchors are used to position virtual references on top of the physical material. Although there is no object tracking involved, the reference will stay aligned with its scrap unless it is moved. \textit{MRUK} additionally handles loading in the scene mesh as a collision surface.

\paragraph{\textbf{Procedural Wood Shader}}
Our interactive wood shader approximates the wood grain with tri-planar mapped voronoi noise. We introduce natural-looking variation to the grain's curvature through a smooth-stepped sine wave function that distorts the noise's linear flow, further stylized with layers of color and simple noise. The grain direction is initialized by rotating the noise around the x-axis. %To ensure that the grain orientation on a 3D part matches its placement on the cut-plan, we introduce additional rotation and position shader parameters. %specifically applied to these parts. These parameters update as the corresponding cut patterns are adjusted.

\section{Case Study}
\label{evaluation}
% \section{Preliminary Evaluation}

\new{We ran a case study with a self-taught woodworker with 10 years of experience designing and building DIY wood assemblies to test \systemName's workflow in practice. Previously interviewed in our formative evaluation (P5), he expressed enthusiasm in exploring the resulting system. The woodworker was unpaid, and we agreed that he  to keep the resulting furniture.} 

\new{The woodworker led the design and fabrication, while an author operated the system. The author used the headset to model designs, incorporating feedback from the woodworker, who viewed the XR feed on a laptop. We chose this setup because the full user interface is not optimized for first-time users. As the system's developer, the author could focus on utilizing all of the features without being distracted by minor technical issues. Acknowledging that this setup deviates from the intended use-case of \systemName's, our goal is not to objectively evaluate the system but to demonstrate its capabilities and technical feasibility.}

%As the system's developer, the author was most familiar with its full functionality and could effectively evaluate all features, instead of relying on the woodworker to discover them within the limited time available. Acknowledging that this setup deviates from \systemName's intended use-case, we instead use this case study to demonstrate its applications for a complex project.}

We aimed to build unique and functional furniture for a university dorm with simple tools (measuring tape, handsaw, and drill) and recycled materials. We asked around in the community to retrieve scraps, and quickly found a construction site that gave us a free pile of discarded wood. 
%To challenge \systemName's ability to enable design for arbitrary, random material, we did not cherry-pick specific pieces, and used the material as it was without any pre-processing. 
The case study spanned two and a half days, including the half a day for sourcing wood. We did not plan what to build in advance, so we were free to improvise based on the available material and the layout of the dorm.

\subsection{Creating the Virtual Inventory}

\begin{figure}[t]
  \centering
  \includegraphics[width=.95\textwidth]{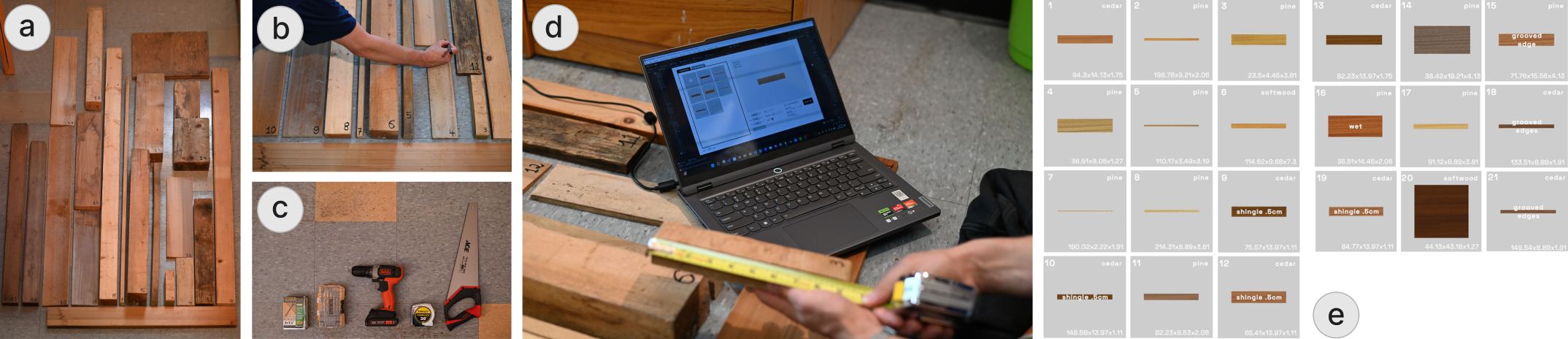}
  \caption{(a) 21 scraps we set aside for our virtual inventory. (b) Numbering the scraps. (c) Tools available for fabrication. (d) Registering our scraps in the virtual inventory. (e) Thumbnails of our final inventory entries.}
  \Description{(a) -- (b) --- (c)---}
  \label{fig:vrregistration}
\end{figure}

The first step was to register our material into \systemName's virtual inventory. Out of the 25 pieces of scrap wood given to us, we selected 21 pieces (\cref{fig:vrregistration}a) for registration. \new{The woodworker} used a marker to number each piece, and input its dimensions, material type, and approximate color into our registration interface using a laptop (\cref{fig:vrregistration}d). There were eight scraps that had considerable imperfections (mostly warping or tapering). These entries were “tagged” with a note so we could look out for them during the design. The registration took about an hour to complete. Although time-consuming, physically handling each scrap to record its details provided us a \textit{\textbf{full tactile understanding}} of the material.
%By physically handling each scrap to record its details, we became aware of its potential and limitations. 
For example, scrap \#16 looked normal, but was wet, whereas scrap \#11 looked rotten, but was unexpectedly sturdy. Throughout this process, we imagined specific functions for each scrap. %, giving us the context necessary to consider what we wanted to build.
Based on these material interactions, the woodworker suggested we build a chair and a smaller, undetermined design. We proceeded with the chair.

\subsection{Chair}

\begin{figure}[h]
  \centering
  \includegraphics[width=.8\textwidth]{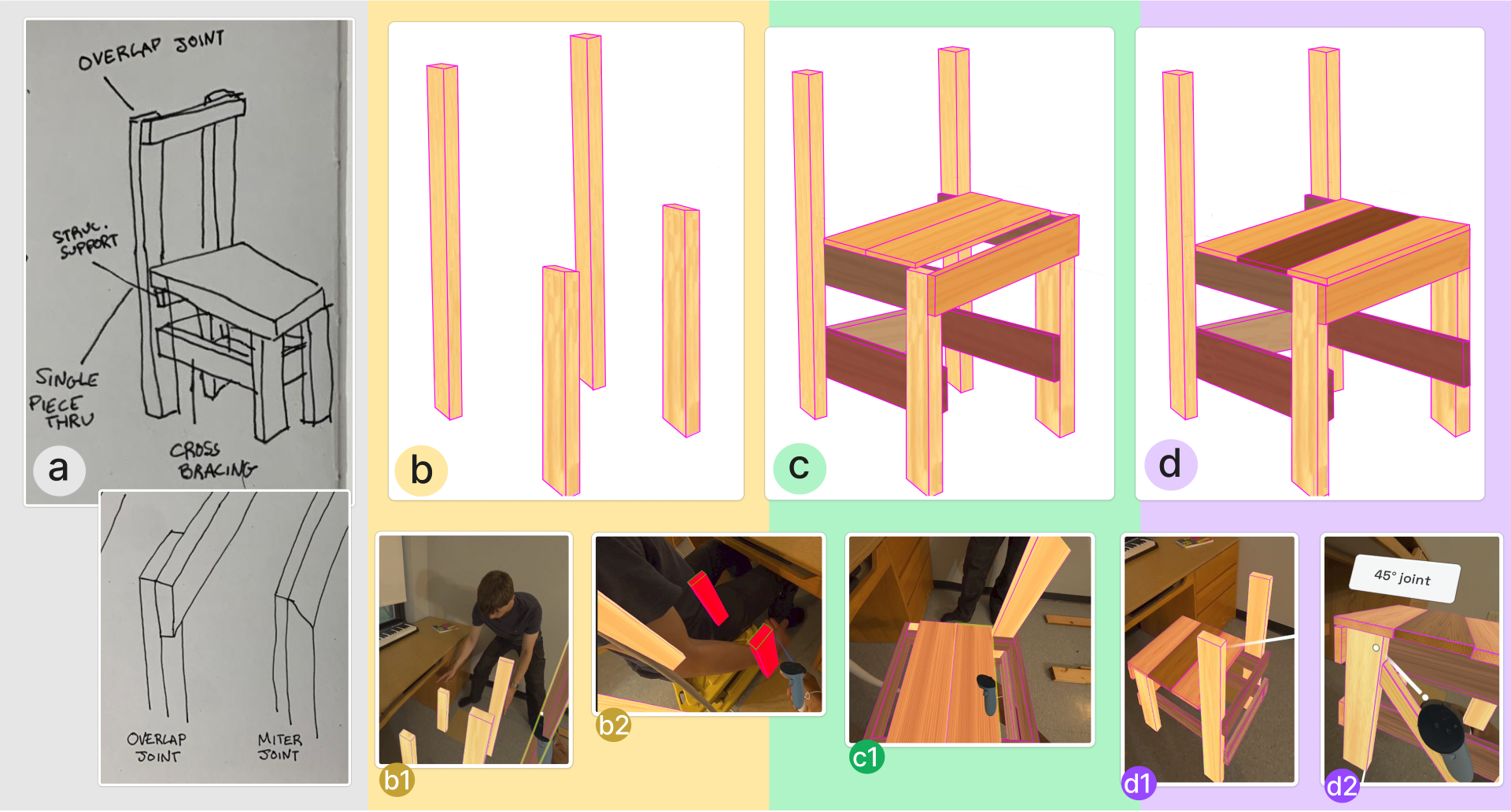}
  \caption{(a) Sketches created by the woodworker while registering wood. (b) Original frame of our chair, which we (b1) designed in situ using (b2) \textit{linked} parts, highlighted red. (c, c1) Framed design could only accommodate two planks with an undesirable gap. (d) We lengthened the seat to fit three planks, fixing the gap. (d2) We also added extra reinforcement.}
  \Description{(a) -- (b) --- (c)---}
  \label{fig:chair}
\end{figure}

Before starting to design in XR, the author used the \textit{Quest 3}'s scene setup to scan the dorm room. %The initial scan failed to capture some irregular wall shapes, but \textit{Meta}'s in-app editing tools allowed us to add the missing surfaces. 
It took 5 minutes to create and clean the scan. The scan was automatically loaded into \systemName's XR environment, along with our saved wood inventory. \new{The woodworker} started to sketch a potential design on paper (\cref{fig:chair}a) during the registration step, but we were unsure how it would map to our scraps. He recommended we build the frame of the chair using our sturdiest wood: scraps \#8 and \#17, and the author used XR to confirm that these pieces could be cut to yield enough material. To ensure the frame fit his body, \new{the woodworker} sat on two stacked tubs, and \new{the author} measured the virtual twins to a comfortable height using \textit{linked} groups (\cref{fig:chair}b2).
%To create the seat surface, we considered two same-width scraps. 
Our original frame only fit a seat made of two equal-width planks side-by-side, leaving an awkward gap at the front (\cref{fig:chair}c1). Instead of cutting a third plank across its length, which is tedious with a handsaw, \new{the author} pulled the frame forward, lengthening the seat to fit three full-width planks (\cref{fig:chair}d1).

\begin{figure}[t]
  \centering
  \includegraphics[width=.8\textwidth]{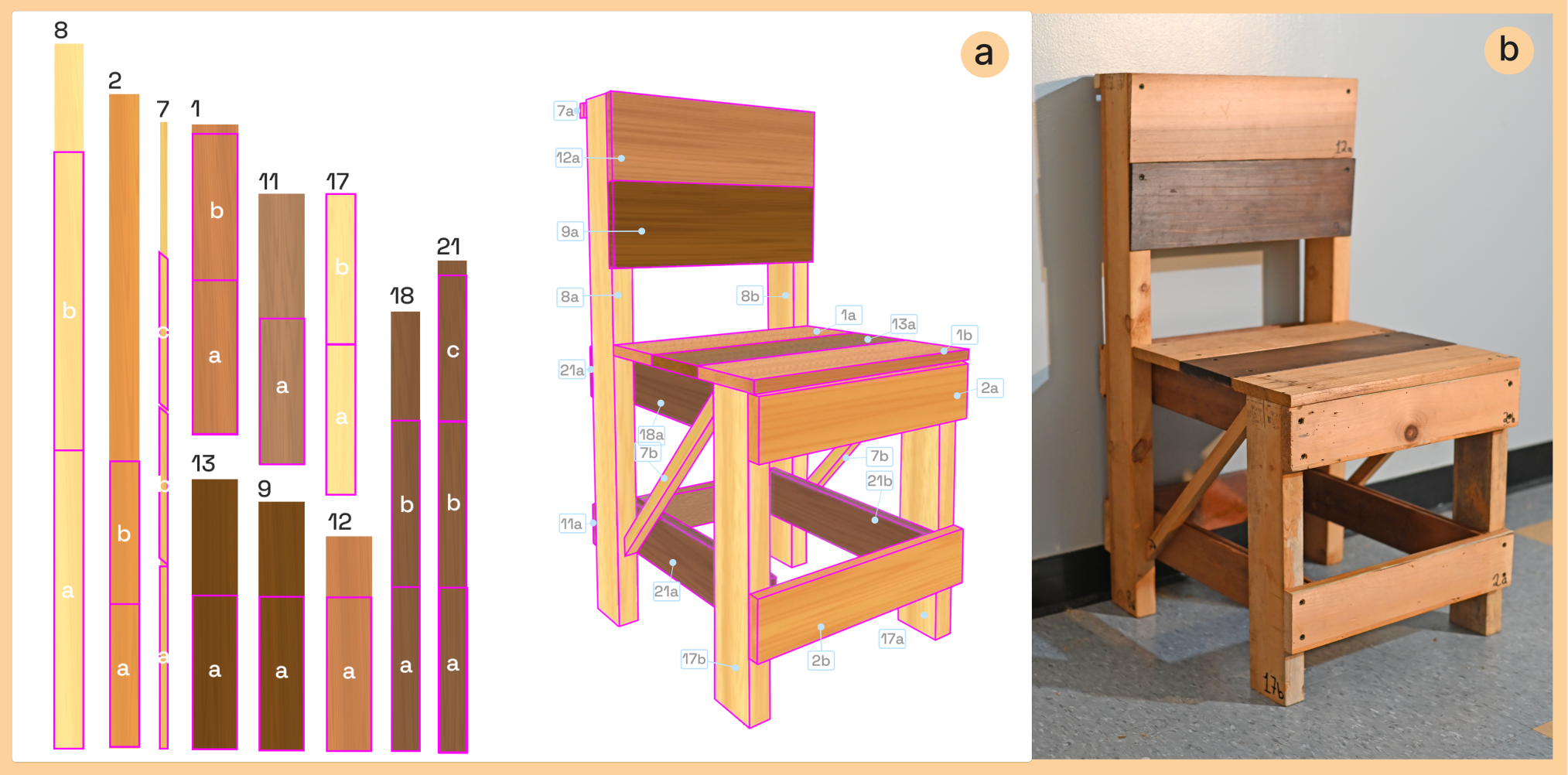}
  \caption{(a) Final cut-plan and spatial model. (b) Fabricated chair.}
  \Description{(a) -- (b) --- (c)---}
  \label{fig:chairfinal}
\end{figure}

 \new{The woodworker guided the author} in adding reinforcements and a backrest to finalize the design. The reinforcements were cut at $45^\circ$ via snapping provided by the system (\cref{fig:chair}d2). The author assigned scraps \#12 and \#9 to the backrest because they were tagged as “shingled”, and the woodworker wanted lighter planks to balance out the heavy back posts. %\new{The woodworker} vertically stacked the physical pieces in the same configuration of our proposed design to verify that they would make a comfortable surface despite their tapered widths. 
\new{The author} used \systemName to split their lengths relative to the chair, and we concluded the 50 minutes XR design session. \new{The woodworker} referenced the final cut plans generated with the system (\cref{fig:chairfinal}a) on \systemName's laptop interface. The built chair took 4 hours to fabricate and is appropriately sized for the woodworker's  body~(\cref{fig:chairfinal}).

\subsection{Slant Shelf}

We also wanted to test the ability of \systemName to assist with a relatively simpler design, so we decided to create a shelving unit. \new{The woodworker} wanted the shelf to be light, involving thin pieces of wood that are easy to cut. There were no such pieces among the leftovers from the chair, so he registered three new scraps to the existing inventory \new{that persisted from our last session.}
\new{The woodworker} discarded these scraps earlier because they were tapered and/or grooved, but they provided the length needed for a taller structure. 
%Out of these new additions, scrap \#22, was particularly unique because of its tapered tip. 
As he handled the new scraps, \new{the woodworker} realized \#22 fit perfectly into a small, odd corner that jutted out of a wall of the dorm, and leaned scrap \#23 against it to create a lean-to structure. This discovery happened outside of XR, but it wasn't obvious how to develop the rest of the design with the few scraps we had left, so we returned to \systemName.

% \new{Once in XR, the author} positioned virtual twins into the same lean-to configuration in the corner, which were captured in the scene mesh. To reinforce the structure, the author's XR experimentation lead to a triangle-shape that could be wedged neatly between the frame. They used \systemName's snapping mechanisms to calculate the its angle.

\begin{figure}[t]
  \centering
  \includegraphics[width=.8\textwidth]{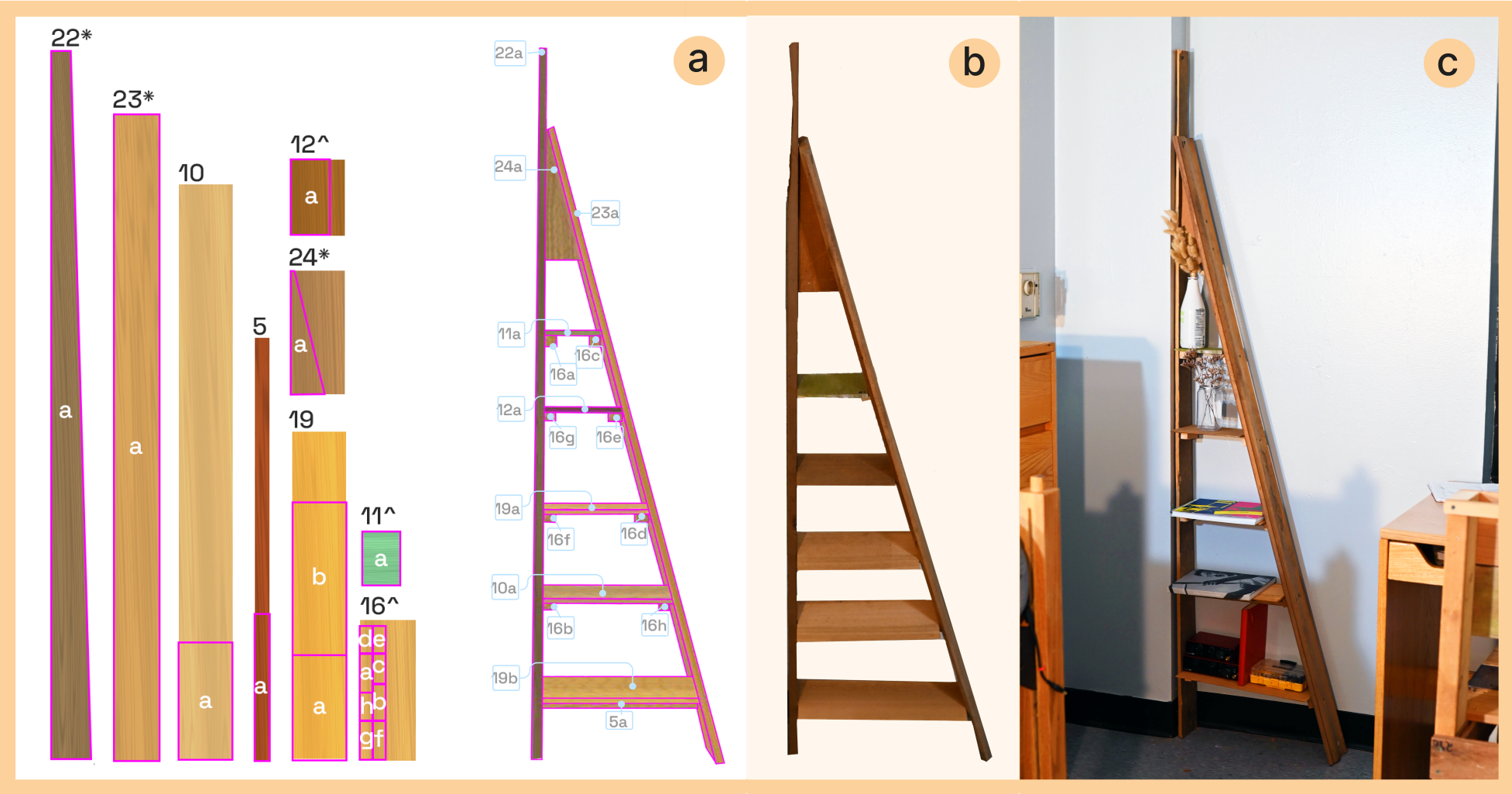}
  \caption{(a) Final cut-plan and spatial model. Pieces numbered with an * were added to the inventory after building the chair. Pieces with a \^{} are cut-offs from the chair. The rest of the pieces were in the original inventory but unused. (b) Photograph of the built shelf. (c) Built shelf in situ.}
  \Description{(a) -- (b) --- (c)---}
  \label{fig:shelf}
\end{figure}

\new{Once in XR, the author} positioned virtual twins into the same lean-to configuration using the scene mesh and cut out a triangle that fit in the tip for reinforcement. They then brought the remaining cut-offs into the scene as shelf candidates, ordered parts by the tapering widths of the frame, and let the collisions between the two framing scraps push them into a flush position. During this process, the inventory motivated us to feature some of our unique material. For example, the backside of scrap \#11 was bright green, likely from algae. The woodworker thought this was appealing, and re-registered its color as green, leading to it being featured as our top shelf. The XR design session lasted 20 minutes and \cref{fig:shelf}a shows our generated cut plans. The built shelf took 1.5 hours to fabricate, fits into the odd corner and can hold a variety of items (\cref{fig:shelf}c).

% . For fabrication, \new{the woodworker} referenced \systemName's design interface to view the cuts and their positions in the shelving unit.

 %This piece was utilized as-is, without a cut, for the top-most shelf.$
% \new{With multiple candidate shelves whose vertical positions were determined by their lengths and therefore our goal of minimizing cuts, previewing variations in XR helped us rapidly iterate towards a functional configuration.}

%and feature some of our unique material. For example, the backside of scrap \#11 was bright green, likely from algae. We thought this was visually-appealing, and re-registered the cut-offs color to green in our inventory. %This piece was utilized as-is, without a cut, for the top-most shelf.$ 
%Finally, \new{the author} placed extra reinforcement bars below each shelf, ending 

\subsection{Material Usage}
Our final virtual inventory contained 24 scraps. 31.08\% of this wood was used for the chair, and 15.22\% for the shelf, totaling 46.03\% of our original scraps. \new{We deliberately created cut plans that preserved larger scraps, so six of our left-over scraps had a surface area larger than 0.15 $m^2$, making them substantial material for future projects}. The smaller off-cuts can still be creatively recycled with our system. 

We found that \systemName's spatial modeling and anchoring systems were precise enough to determine material usage in advance. As the woodworker fabricated the designs, he used the dimensions from the cut plan for the \new{frame of the chair and the shelves of the shelving unit,} referencing the partial assemblies to determine the remaining cuts. He found this to be more intuitive, because parts naturally shift slightly during fabrication. As predicted by our system, we had sufficient material to cut all parts. \new{The average deviation between the cut plan and fabricated lengths was 7 mm for the chair and 3 mm for the shelving unit. The deviation for the chair was higher because its framing parts were imperfectly measured.} %Still, we had enough material to create all of the cuts proposed by \systemName's cut-plan, which meant our virtual cuts accurately reflected available dimensions.

\subsection{Reflection}

\begin{figure}[t]
  \centering
  \includegraphics[width=.85\textwidth]{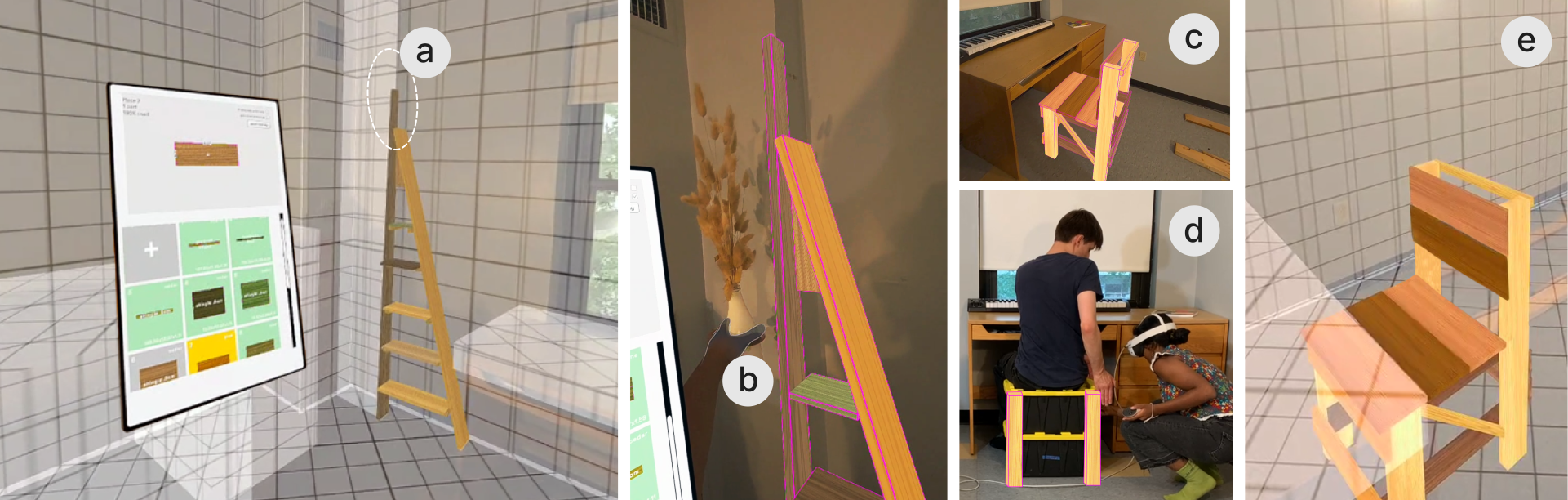}
  \caption{\systemName enabled in situ decision making in several ways: (a) Designing the shelf to fit an impractical corner. (b) Checking if a vase fits on the shelf. (c) Checking if a desk fits with the chair. (d) Designing the chair around the woodworker’s body. (e) Determining the chair’s height using the floor plane of the scene mesh.}
  \Description{(a) -- (b) --- (c)---}
  \label{fig:situ}
\end{figure}

A key strength of our system in practice was that it facilitated \new{decision making \textit{in situ}}. Even in a standard sized dorm room, we frequently used the scene mesh for alignment (\cref{fig:situ}a,e), and the visual context of the surroundings to inform design decisions (\cref{fig:situ}b,c,d). \new{For example, when designing the chair, the author could quickly assemble a frame suited to the woodworker’s body and obtain immediate feedback from him. The woodworker later shared that \textit{"using actual objects and spaces themselves to measure things, as opposed to taking a tape-measure and then manually drawing it... This really suited the way I work."}}  

 %(f) The physical wood next to its virtual twin.
 %We can also confirm that the headset's camera resolution was sufficient for the author to engage with both the woodworker and physical materials (\cref{fig:situ}f).

Coupling the virtual inventory with physical scraps helped us identify specific uses for varied materials that otherwise proved difficult to visualize. \new{Even after the author defined the chair's rough proportions in situ, they were not sure if they had enough material for its surfaces, so they relied on 
%spatial context gained by handling and cutting virtual twins.  
 material usage indicators and linked editing to determine which scraps to cut from. 
%By handling and virtually "cutting" the planks in space, the author received spatial context similar to physically moving the wood,  while the system’s 3D collision detection and linked editing tools provided the control required for this process. 
When designing the shelf, finding a practical shelving arrangement that minimized cuts was our biggest challenge because each shelf position depended on its length. \systemName allowed us to preview all shelves at once and iterate towards a functional layout. The woodworker felt that \systemName's inventory complemented the exploratory nature of his practice: \textit{“This system is faster than anything I’ve used to visualize what I can make with scraps... It feels like the scraps are this finite resource we can freely allocate...and play with."}}

An unexpected affordance of our system was our ability to collaborate within it, \new{as \systemName was originally designed for a single user with woodworking knowledge. The author did not have prior experience with woodwork design, but prototyping in \systemName allowed them to express design intent in a way that the woodworker could understand and guide.} A major limitation observed by the woodworker was the need to repeatedly travel between the physical material (to check the physical cut overlays) and the 3D assembly. Since the scraps we worked with lacked localized surface features, \new{we generated cut plans in \textit{auto-resolve} mode and accepted the compact packing configurations.} However, if our scraps had specific features to design around, this approach would not have been sufficient.

% This setup allowed the woodworker to focus on evaluating the design and providing feedback, not just verbally, but \textit{spatially}, turning his hands and physical material into reference points.

%Even if we were willing to continuously reference the cut-overlays, it would be difficult to extrapolate any marks or imperfections back onto the spatial model. 

%\begin{figure}[h]
%  \centering
%  \includegraphics[width=.5\textwidth]{figures/in-context.png}
%  \caption{The final chair and shelving unit together. The shelving unit can also function as a light stand.}
%  \Description{(a) -- (b) --- (c)---}
%  \label{fig:finalsitu}
%\end{figure}

\section{Feedback From Target Users}
\new{We held feedback sessions with eight (U1-U8) hobbyist woodworkers, ages 21-40, to learn how they might feel about engaging with \systemName. All participants had experience building functional items with scrap wood, had used an XR headset at least once, and were recruited as unpaid volunteers from the local community. As indicated by our case study, the design and fabrication of a full piece of furniture is a time and labor-intensive process. Meanwhile, user-testing an isolated feature would not capture the full experience of creating a situated, material-aware design. \finaledit{Therefore, we presented a video including the system's motivation, walk-through and case study results (refer to the video figure)} inspired by the evaluation approach of \textit{Hybrid Carpentry} \cite{zoran_human-computer_2013}, which was followed up with a short, open-ended XR design session so participants could try the system themselves.}

\begin{figure}[h]
  \centering
  \includegraphics[width=.95\textwidth]{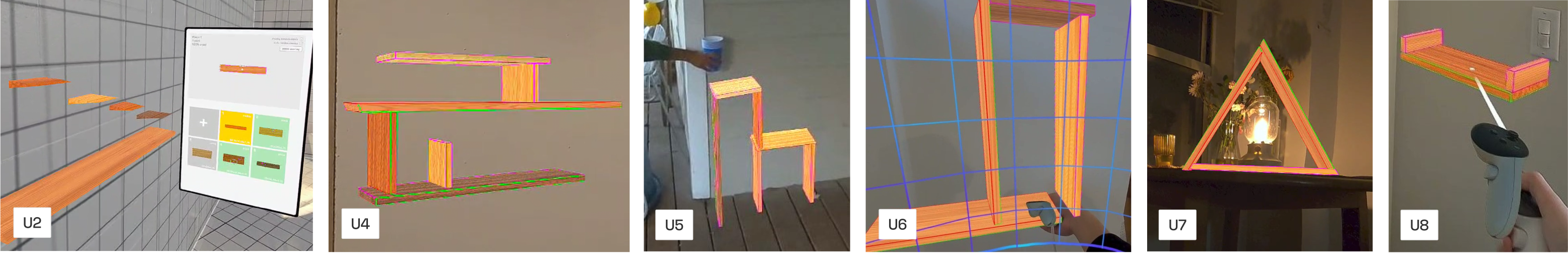}
  \caption{Select simple shelves designed by feedback session participants during their 5-minute free-design period.}
  \Description {a,b,c}
  \label{fig:evaluation}
\end{figure}

\new{Each session lasted 1 hour, with 45 minutes dedicated to semi-structured discussion, including watching the video, and 15 minutes for hands-on use. The XR portion started with a 10 minute author-guided tutorial covering core \systemName mechanisms: accessing the inventory, positioning and cutting parts in 3D, duplicating parts for linked edits and re-assigning parts. Participants spent their last 5 minutes with \systemName freely designing a simple shelving unit (\cref{fig:evaluation}) using five pre-registered scraps, giving us oral feedback both during and after headset use. We share our findings below.} %The author verbally reminded the participant of any controller mappings they may have forgotten, but did not intervene otherwise.}

\subsection{Usability of the XR Interface}
% \paragraph{\textbf{Usability of the XR Interface}}
\new{Every participant could grasp the \systemName's XR modeling interface during the 10 minute tutorial. U4 said our operations to \textit{“cut”} wood felt \textit{“really straightforward...you just need to select the edges you want to push in or push out”.} U2 thought that the cut operations \textit{“felt very physical”}. U6 shared, \textit{"[I like] that I can actually see the [virtual] wood in real life, it's like I'm building it! It feels easier than CAD."} U2, U4, U5, and U7 felt that \systemName's spatial interface could help them explore their design space in an adaptive way. U2 shared, \textit{“I love the creativity and the free flow of placing [virtual] wood wherever I want,”} as she usually gets "\textit{stuck}" thinking about \textit{“L-brackets and stuff.”}}
\new{However, U1 felt that placing each part in space by hand felt \textit{“tedious”} and said he would strongly prefer to adjust pre-defined designs, rather than creating a design from scratch. We also observed limitations with our current spatial modeling mechanisms: all participants became frustrated by 3D parts unexpectedly \textit{"floating away"} because of our 3D collision handling and experienced difficulty aligning them in flush arrangements. U7 also shared that she wouldn't feel confident designing complex objects in XR (referencing the chair from the case study) without "\textit{some sort of a guide,}" such as a sample 3D model to reference.}

\subsection{Designing In Situ}
% \paragraph{\textbf{Designing In Situ}}
\new{All participants valued \systemName's support for designing in context. U1 thought \systemName "\textit{is most useful for environment specific projects, [like] a shelf that fits an exact space.}”  U3 reflected on past experiences designing and fabricating furniture that did not quite “\textit{fit}” in its intended space, and felt that \systemName could help him prevent such issues. U1, U2, U4, and U6, with experience using XR to preview furniture/product designs, appreciated the ability to \textit{edit} their design in context, a feature lacking in the systems they had previously used. During our XR demos, each participant could successfully align their shelves with the scene mesh and felt that it helped them verify fit. However, U5 suggested using a more transparent texture for the scene mesh to improve visibility of the environment behind it. After trying our system, U3 expressed interest in using it to mock-up a wooden stand he was planning to build "\textit{just to see how it would look and fit in the space}”. U5, U6, and U8 indicated they would want to use \systemName for larger, environment-specific projects rather than small builds. U5 felt that \textit{“[getting out an XR headset] might not be worth the effort”} for a small, standalone object.}

\subsection{Designing with Scraps}
% \paragraph{\textbf{Designing with Scraps}}
\new{U5, who builds functional structures for her farm, noted that \systemName's scrap inventory could \textit{"increase [her] creativity"} while motivating her to be \textit{"sustainable"}. Participants generally felt that \systemName's feedback features, such as red indicators for material limits, could ensure their designs remain feasible. U4 shared, "\textit{I like the blocks turning red because it lets me know how much material I still have left}," as he edited a part in space. U2 felt like she could experiment in XR instead of “\textit{wasting a real piece of wood,}” by letting the 2D cut plan automatically generate as she worked.} \new{However, U3 found it difficult to follow “\textit{the relationship between the stock and cut part}” because they were “\textit{changing both in space and on the [2D] menu}”. He explained that he does not think about his 3D designs in terms of 2D parts until it is time to acquire material.}
\new{Although participants appreciated the display of grain and color on the scraps, nearly everyone expressed interest in using scanned textures instead of initializing this information themselves. The only exception was U4, who preferred the simulated textures for their “\textit{simplicity}.”}

\subsection{Reflection}
% \paragraph{\textbf{Target Audience}}
\new{We designed \systemName to address challenges faced by "DIY woodworkers," recognizing that this group encompasses a broad range of approaches. Our feedback sessions helped verify our target audience: U2, U4, U5 and U7, who build quick, functional projects from scraps, liked \systemName's material-aware features the most. In contrast, U1 and U3, who work on more sophisticated, time-consuming builds, expressed a preference for purchasing new materials for a predefined design rather than adapting the design to fit available scraps. They were understandably skeptical of our system's scrap inventory. While U6 and U8 enjoy working with scraps for detailed wood craft, they found \systemName's design space too limited for their needs. Therefore, we can conclude that \systemName is most useful for DIY woodworkers who prioritize material-driven, \textit{adaptive} workflows over intricate craftsmanship. This group are not necessarily novices; they are simply less outcome-oriented in their approach.}

\section{Discussion}
% \hl{[working on this section]} 
% In this section, we discuss the limitations of \systemName~ and potential directions for future research.

% \subsection{Limitations and Future Work}
\new{In this section, we discuss findings from our case study and feedback sessions, and present transferable principles for designing systems that aim to consider material and/or space in a similar way.}

\paragraph{\textbf{Interactive Design vs. Fabrication for Scraps.}}
\new{Currently, \systemName supports interactive \textit{design}, but not fabrication. We assume that users will finalize their design before beginning fabrication, with adjustments only made to address unexpected errors during the build. Furthermore, supporting fabrication in XR would be unwieldy and potentially dangerous with our current headset. However, we acknowledge that working with scraps is an inherently nonlinear process. \systemName addresses this by encouraging users to explore their full design space \textit{before} making physical cuts, reducing the likelihood of costly mistakes while supporting the improvisation necessitated by scraps. In general, similar material-aware systems should prioritize iterative exploration during the design phase, providing users with tools to visualize, manipulate, and “test” their designs before moving to physical fabrication. At the same time, integrating some features that accommodate mid-fabrication adjustments—such as material re-assignment or pattern editing—can ensure the system meets the reality of manual workflows.}

\paragraph{\textbf{Digital Design with Non-Standard Materials.}}
\new{By virtue of being in XR, our system allows users to engage with real wood during their design process. We expect users to naturally reference both the virtual inventory and their \textbf{physical inventory} when selecting suitable scraps for their design, which is in line with the observation that tactile feedback is essential for raw material \textit{crafts} \cite{crevels_coarse_2023}. We ensured that all spatial design operations outlined can be performed with a single controller, so the other hand is free to interface with physical material (we had previously considered enabling the headset's built in hand-tracking, but it proved too imprecise in its current state). Based on feedback from formative interviews, we chose user-guided layout creation over automated material optimization, as this approach better accommodates the variability of natural material. During our case study, both the woodworker and author relied on physical interactions with material to assign specific scraps to different parts of our designs. If the system had automatically made all assignments, the resulting designs would not have met our structural or aesthetic vision.}

\paragraph{\textbf{Creative Design in Situ.}}
\new{From our formative interviews, we identified a need for systems that help woodworkers visualize their designs in context, and this was unanimously reaffirmed in our feedback sessions. Furthermore, these sessions helped us formulate a new insight: participants found value in the ability to \textit{create} designs directly in situ. U1, for instance, had previously tried an AR CAD plugin to preview his woodwork designs, but disliked having to switch between AR and his computer every time he made an adjustment. He appreciated that \systemName allows users to \textit{“directly adjust things in place.”} While in situ design systems naturally embed spatial context into the design process, we observe that they benefit from additional constraints: since \systemName does not utilize parametric templates, features such as material usage limits and environment snapping are necessary to guide the design process. Our feedback sessions revealed that users would appreciate even more constraints, such as structural validation. We also found that supporting a simple set of design operations (instead of the complexity of commercial CAD) means designs created with our system can be fabricated with basic hand/power tools and are easier to manipulate in XR.}

\section{Future Work}
% \new{We outline opportunities for future work based on findings from our case study and feedback sessions.}

\paragraph{\textbf{Multi-User Scenarios.}}
\new{Two of our participants from our feedback sessions, U2 and U3, are a couple who team up on woodworking projects. After seeing the case study’s two-person setup, they expressed a desire in having two headsets for collaboration, so they could develop their designs in shared spatial context. However, they suggested that only one person operates the system while the other directs, so the non-headset wearer can focus on giving feedback and preparing materials. Future work might explore both multi-headset and asymmetric setups for co-design in XR.} %These roles match those taken on in our case study, where the author was the operator as the woodworker directed them. The woodworker also intermittently wore the headset to assess spatial occupancy, corroborating the value of a multi-headset setup.} %Alternatively, integrating a more robust asymmetric setup for collaboration has its own value, as outlined by \textit{TransceiVR} \cite{thoravi_kumaravel_transceivr_2020}.

\paragraph{\textbf{Scanned Inventory.}}
After challenges visualizing localized imperfections during our case study and feedback from most participants that manual registration seemed tedious, we consider integrating a 3D scanning workflow as a promising next step (\textit{Meta} plans to provide developer \textit{Passthrough} camera access in early 2025\footnote{\textit{Meta 2024 Developer Keynote} \url{https://developers.facebook.com/m/meta-connect-developer-sessions/developer-keynote}}). 
% \finaledit{We envision that materials could also be labeled with visible~\cite{dogan_structcode_2023, campos_zamora_moirewidgets_2024} or unobtrusive~\cite{dogan_brightmarker_2023, dogan_infraredtags_2022, dogan_standarone_2023} markers to facilitate automatic machine registration.}
However, we acknowledge that an important part of woodworking involves the physical inspection and engagement of materials \cite{crevels_coarse_2023}. Therefore, a future iteration of \systemName can explore a hybrid approach combining the convenience of a scanned inventory with tactile material interactions. For example, a woodworker can optionally "tag" scanned features with more detail after manually inspecting their material.
% Emerging AI tools, such as multimodal large language models, can also be helpful in predicting or suggesting material types and imperfections~\cite{dogan_augmented_2024, dogan_ubiquitous_2024}.}

\new{Currently, scrap registration in \systemName occurs on the laptop because it is difficult to read a tape measure at the \textit{Quest's} current \textit{Passthrough} resolution. We acknowledge that switching between the headset and laptop breaks flow, and we anticipate in-headset registration to become viable as camera technologies advance, perhaps aided by marker-based tracking \cite{lau_situated_2012, jahn_mixed_2024} or gestural localization \cite{htet_kyaw_gesture_2024}.}

\paragraph{\textbf{Expanded Modeling Utilities.}}
\new{Our system only creates 8-cornered rectilinear shapes, but U1, U6 and U8 expressed interest in support for additional joints and cuts, such as \textit{dados}, \textit{rabbets}, and \textit{grooves}.} %\new{Future work could also integrate a library of standard connector parts, such as L-brackets, screws, and dowels. A user could use these parts to verify that their design has enough tolerance to accommodate necessary connectors. 
\new{Six of our feedback session participants also expressed a strong interest in having access to "templates" for standard structures, such as chairs, stairs and tables.} This could expand to integrating the situated parametric design features found in \textit{pARam} \cite{stemasov_param_2024} with the material-awareness of \systemName. \new{For instance, a chair "template" could allow users to adjust parameters such as seat height or backrest angle, while the system prompts the user to assign scraps to the design.}
%Supporting slightly more complex geometry would also allow users to explore new creative designs, such as a 5-sided front piece for a birdhouse. We are also interested in exploring additional utilities for repeating parts such as: distributing parts with consistent spacing or aligning parts to a specified axis. 
% We see these templates integrating with the system’s in-situ and material-aware qualities, such as automatically adjusting shelf spacing to ceiling height or suggesting scrap assignments for unassigned parts.

\paragraph{\textbf{Structural Design Guidance.}}
Although our interface is geared towards casual makers, it necessitates some tacit woodworking knowledge. For example, the woodworker we collaborated in the case-study was experienced in constructing stable wood assemblies, and knew how to correctly reinforce our virtual designs. In contrast, completely inexperienced makers may not know how to create a sturdy structure. \new{Five of our feedback session participants shared that they wanted the system to tell them if their designs were structurally sound. Accordingly, we could provide feedback on physical validity, as seen in previous works \cite{umetani_guided_2012, saul_sketchchair_2010}. Integrating these approaches with our XR system could enable more interactive evaluations, allowing users to 'push' on a virtual table or 'sit' on a chair to simulate how their designs might respond.}

% The current setup requires some prior knowledge of both XR systems and woodworking, which could be a barrier for novices. Providing more intuitive onboarding experiences and tutorials within the system could help bridge this gap.

% The potential of \systemName extends beyond its current capabilities. Future research could explore integrating more sophisticated design optimization algorithms that account for both material constraints and aesthetic goals, thus enabling users to achieve more refined designs with minimal waste. Another exciting avenue is the incorporation of collaborative features that allow multiple users to design and refine projects together in a shared XR environment, promoting community-driven innovation in sustainable woodworking.
\section{Conclusion}

In this paper, we introduced \systemName, an XR system designed to support material-aware, in situ design for DIY woodworking projects. \systemName enables users to plan and execute designs that are within their functional reach. Our case study demonstrated the ability of our proposed system to facilitate design with irregular materials, highlighting its potential to bridge the gap between digital planning and manual assembly for casual makers. Our user feedback sessions show that DIY woodworkers appreciated the system's support of creative, situated design with scrap materials and helped verify the target audience for such tools. By contributing a digital tool that enhances the way woodworkers naturally approach projects, we can enable them to experiment and plan more effectively without losing the tactile, hands-on experience of their craft.

\begin{acks}
We extend special thanks to woodworker Geoffrey Hazard, whose skill and insight were invaluable in demonstrating our system via case study. We also thank everyone who generously volunteered their time for our interviews and feedback sessions, as well as Bo Zhu and Haoran Xie for their ongoing support. This research was supported by the NSF IRES program (Award: 2433313), JST ACT-X (Award: JPMJAX210P), JST AdCORP (Award: JPMJKB2302), JSPS KAKENHI (Award: JP23K19994) and a collaborative research fund between Mercari Inc. R4D and RIISE, University of Tokyo.

\end{acks}

\bibliographystyle{ACM-Reference-Format}
\bibliography{bib_woodwork}


\begin{thebibliography}{69}

%%% the Web address.

\ifx \showCODEN    \undefined \def \showCODEN     #1{\unskip}     \fi
\ifx \showDOI      \undefined \def \showDOI       #1{#1}\fi
\ifx \showISBNx    \undefined \def \showISBNx     #1{\unskip}     \fi
\ifx \showISBNxiii \undefined \def \showISBNxiii  #1{\unskip}     \fi
\ifx \showISSN     \undefined \def \showISSN      #1{\unskip}     \fi
\ifx \showLCCN     \undefined \def \showLCCN      #1{\unskip}     \fi
\ifx \shownote     \undefined \def \shownote      #1{#1}          \fi
\ifx \showarticletitle \undefined \def \showarticletitle #1{#1}   \fi
\ifx \showURL      \undefined \def \showURL       {\relax}        \fi
% invisible to TeX
\providecommand\bibfield[2]{#2}
\providecommand\bibinfo[2]{#2}
\providecommand\natexlab[1]{#1}
\providecommand\showeprint[2][]{arXiv:#2}

\bibitem[Agrawal et~al\mbox{.}(2015)]%
        {agrawal_protopiper_2015}
\bibfield{author}{\bibinfo{person}{Harshit Agrawal}, \bibinfo{person}{Udayan Umapathi}, \bibinfo{person}{Robert Kovacs}, \bibinfo{person}{Johannes Frohnhofen}, \bibinfo{person}{Hsiang-Ting Chen}, \bibinfo{person}{Stefanie Mueller}, {and} \bibinfo{person}{Patrick Baudisch}.} \bibinfo{year}{2015}\natexlab{}.
\newblock \showarticletitle{Protopiper: {Physically} {Sketching} {Room}-{Sized} {Objects} at {Actual} {Scale}}. In \bibinfo{booktitle}{\emph{Proceedings of the 28th {Annual} {ACM} {Symposium} on {User} {Interface} {Software} \& {Technology}}}. \bibinfo{publisher}{ACM}, \bibinfo{address}{Charlotte NC USA}, \bibinfo{pages}{427--436}.
\newblock
\showISBNx{978-1-4503-3779-3}
\urldef\tempurl%
\url{https://doi.org/10.1145/2807442.2807505}
\showDOI{\tempurl}


\bibitem[Arisandi et~al\mbox{.}(2014)]%
        {arisandi_virtual_2014}
\bibfield{author}{\bibinfo{person}{Ryan Arisandi}, \bibinfo{person}{Mai Otsuki}, \bibinfo{person}{Asako Kimura}, \bibinfo{person}{Fumihisa Shibata}, {and} \bibinfo{person}{Hideyuki Tamura}.} \bibinfo{year}{2014}\natexlab{}.
\newblock \showarticletitle{Virtual {Handcrafting}: {Building} {Virtual} {Wood} {Models} {Using} {ToolDevice}}.
\newblock \bibinfo{journal}{\emph{Proc. IEEE}} \bibinfo{volume}{102}, \bibinfo{number}{2} (\bibinfo{date}{Feb.} \bibinfo{year}{2014}), \bibinfo{pages}{185--195}.
\newblock
\showISSN{1558-2256}
\urldef\tempurl%
\url{https://doi.org/10.1109/JPROC.2013.2294243}
\showDOI{\tempurl}
\newblock
\shownote{Conference Name: Proceedings of the IEEE}.


\bibitem[Baudisch et~al\mbox{.}(2019)]%
        {baudisch_kyub_2019}
\bibfield{author}{\bibinfo{person}{Patrick Baudisch}, \bibinfo{person}{Arthur Silber}, \bibinfo{person}{Yannis Kommana}, \bibinfo{person}{Milan Gruner}, \bibinfo{person}{Ludwig Wall}, \bibinfo{person}{Kevin Reuss}, \bibinfo{person}{Lukas Heilman}, \bibinfo{person}{Robert Kovacs}, \bibinfo{person}{Daniel Rechlitz}, {and} \bibinfo{person}{Thijs Roumen}.} \bibinfo{year}{2019}\natexlab{}.
\newblock \showarticletitle{Kyub: {A} {3D} {Editor} for {Modeling} {Sturdy} {Laser}-{Cut} {Objects}}. In \bibinfo{booktitle}{\emph{Proceedings of the 2019 {CHI} {Conference} on {Human} {Factors} in {Computing} {Systems}}}. \bibinfo{publisher}{ACM}, \bibinfo{address}{Glasgow Scotland Uk}, \bibinfo{pages}{1--12}.
\newblock
\showISBNx{978-1-4503-5970-2}
\urldef\tempurl%
\url{https://doi.org/10.1145/3290605.3300796}
\showDOI{\tempurl}


\bibitem[Beltagui et~al\mbox{.}(2021)]%
        {beltagui_bricolage_2021}
\bibfield{author}{\bibinfo{person}{Ahmad Beltagui}, \bibinfo{person}{Achilleas Sesis}, {and} \bibinfo{person}{Nikolaos Stylos}.} \bibinfo{year}{2021}\natexlab{}.
\newblock \showarticletitle{A bricolage perspective on democratising innovation: {The} case of {3D} printing in makerspaces}.
\newblock \bibinfo{journal}{\emph{Technological Forecasting and Social Change}}  \bibinfo{volume}{163} (\bibinfo{date}{Feb.} \bibinfo{year}{2021}), \bibinfo{pages}{120453}.
\newblock
\showISSN{0040-1625}
\urldef\tempurl%
\url{https://doi.org/10.1016/j.techfore.2020.120453}
\showDOI{\tempurl}


\bibitem[Besserer et~al\mbox{.}(2021)]%
        {besserer_cascading_2021}
\bibfield{author}{\bibinfo{person}{Arnaud Besserer}, \bibinfo{person}{Sarah Troilo}, \bibinfo{person}{Pierre Girods}, \bibinfo{person}{Yann Rogaume}, {and} \bibinfo{person}{Nicolas Brosse}.} \bibinfo{year}{2021}\natexlab{}.
\newblock \showarticletitle{Cascading {Recycling} of {Wood} {Waste}: {A} {Review}}.
\newblock \bibinfo{journal}{\emph{Polymers}} \bibinfo{volume}{13}, \bibinfo{number}{11} (\bibinfo{date}{May} \bibinfo{year}{2021}), \bibinfo{pages}{1752}.
\newblock
\showISSN{2073-4360}
\urldef\tempurl%
\url{https://doi.org/10.3390/polym13111752}
\showDOI{\tempurl}


\bibitem[Bezci(2016)]%
        {bezci_it_2016}
\bibfield{author}{\bibinfo{person}{Ismail Bezci}.} \bibinfo{year}{2016}\natexlab{}.
\newblock \showarticletitle{Do {It} {Yourself}: {A} {Methodological} {Study} in {Furniture} {Design}}. In \bibinfo{booktitle}{\emph{3rd {International} {Multidisciplinary} {Scientific} {Conference} on {Social} {Sciences} and {Arts}}}, Vol.~\bibinfo{volume}{16}. \bibinfo{pages}{313--324}.
\newblock
\showISBNx{978-619-7105-78-0}
\urldef\tempurl%
\url{https://doi.org/10.5593/SGEMSOCIAL2016/B43/S15.038}
\showDOI{\tempurl}


\bibitem[Bonvoisin et~al\mbox{.}(2017)]%
        {bonvoisin_design_2017}
\bibfield{author}{\bibinfo{person}{Jérémy Bonvoisin}, \bibinfo{person}{Jahnavi~Krishna Galla}, {and} \bibinfo{person}{Sharon Prendeville}.} \bibinfo{year}{2017}\natexlab{}.
\newblock \showarticletitle{Design {Principles} for {Do}-{It}-{Yourself} {Production}}. In \bibinfo{booktitle}{\emph{Sustainable {Design} and {Manufacturing} 2017}}, \bibfield{editor}{\bibinfo{person}{Giampaolo Campana}, \bibinfo{person}{Robert~J. Howlett}, \bibinfo{person}{Rossi Setchi}, {and} \bibinfo{person}{Barbara Cimatti}} (Eds.). \bibinfo{publisher}{Springer International Publishing}, \bibinfo{address}{Cham}, \bibinfo{pages}{77--86}.
\newblock
\showISBNx{978-3-319-57078-5}
\urldef\tempurl%
\url{https://doi.org/10.1007/978-3-319-57078-5_8}
\showDOI{\tempurl}


\bibitem[Bourgault et~al\mbox{.}(2023)]%
        {bourgault_coilcam_2023}
\bibfield{author}{\bibinfo{person}{Samuelle Bourgault}, \bibinfo{person}{Pilar Wiley}, \bibinfo{person}{Avi Farber}, {and} \bibinfo{person}{Jennifer Jacobs}.} \bibinfo{year}{2023}\natexlab{}.
\newblock \showarticletitle{{CoilCAM}: {Enabling} {Parametric} {Design} for {Clay} {3D} {Printing} {Through} an {Action}-{Oriented} {Toolpath} {Programming} {System}}. In \bibinfo{booktitle}{\emph{Proceedings of the 2023 {CHI} {Conference} on {Human} {Factors} in {Computing} {Systems}}}. \bibinfo{publisher}{ACM}, \bibinfo{address}{Hamburg Germany}, \bibinfo{pages}{1--16}.
\newblock
\showISBNx{978-1-4503-9421-5}
\urldef\tempurl%
\url{https://doi.org/10.1145/3544548.3580745}
\showDOI{\tempurl}


\bibitem[Bunn(1999)]%
        {bunn_importance_1999}
\bibfield{author}{\bibinfo{person}{Stephanie Bunn}.} \bibinfo{year}{1999}\natexlab{}.
\newblock \showarticletitle{The {Importance} of {Materials}}.
\newblock \bibinfo{journal}{\emph{Journal of Museum Ethnography}} \bibinfo{volume}{1}, \bibinfo{number}{11} (\bibinfo{year}{1999}), \bibinfo{pages}{15--28}.
\newblock
\showISSN{0954-7169}
\urldef\tempurl%
\url{https://www.jstor.org/stable/40793620}
\showURL{%
\tempurl}
\newblock
\shownote{Publisher: Museum Ethnographers Group}.


\bibitem[Camburn and Wood(2018)]%
        {camburn_principles_2018}
\bibfield{author}{\bibinfo{person}{Bradley Camburn} {and} \bibinfo{person}{Kristin Wood}.} \bibinfo{year}{2018}\natexlab{}.
\newblock \showarticletitle{Principles of maker and {DIY} fabrication: {Enabling} design prototypes at low cost}.
\newblock \bibinfo{journal}{\emph{Design Studies}}  \bibinfo{volume}{58} (\bibinfo{date}{Sept.} \bibinfo{year}{2018}), \bibinfo{pages}{63--88}.
\newblock
\showISSN{0142-694X}
\urldef\tempurl%
\url{https://doi.org/10.1016/j.destud.2018.04.002}
\showDOI{\tempurl}


\bibitem[Cousin et~al\mbox{.}(2023)]%
        {cousin_wild_2023}
\bibfield{author}{\bibinfo{person}{Tim Cousin}, \bibinfo{person}{Latifa Alkhayat}, \bibinfo{person}{Natalie Pearl}, \bibinfo{person}{Christopher~B. Dewart}, {and} \bibinfo{person}{Caitlin Mueller}.} \bibinfo{year}{2023}\natexlab{}.
\newblock \showarticletitle{Wild {Wood} {Gridshells}: {Mixed}-{Reality} {Construction} of {Nonstandard} {Wood}}.
\newblock \bibinfo{journal}{\emph{Technology{\textbar}Architecture + Design}} \bibinfo{volume}{7}, \bibinfo{number}{2} (\bibinfo{date}{July} \bibinfo{year}{2023}), \bibinfo{pages}{216--231}.
\newblock
\showISSN{2475-1448}
\urldef\tempurl%
\url{https://doi.org/10.1080/24751448.2023.2245725}
\showDOI{\tempurl}
\newblock
\shownote{Publisher: Routledge \_eprint: https://doi.org/10.1080/24751448.2023.2245725}.


\bibitem[Crevels(2023)]%
        {crevels_coarse_2023}
\bibfield{author}{\bibinfo{person}{Eric Crevels}.} \bibinfo{year}{2023}\natexlab{}.
\newblock \showarticletitle{Coarse epistemes: {Skill}, craftsmanship and tacit knowledge in the grit of the world}.
\newblock \bibinfo{journal}{\emph{Perspectives on Tacit Knowledge in Architecture}} (\bibinfo{year}{2023}), \bibinfo{pages}{13--29}.
\newblock


\bibitem[Deshpande et~al\mbox{.}(2022)]%
        {deshpande_upcycling_2022}
\bibfield{author}{\bibinfo{person}{Himani Deshpande}, \bibinfo{person}{Jin Yu}, \bibinfo{person}{Akash Talyan}, {and} \bibinfo{person}{Noah Posner}.} \bibinfo{year}{2022}\natexlab{}.
\newblock \showarticletitle{Upcycling discarded {HDPE} plastic bags for creative exploration in product design}. In \bibinfo{booktitle}{\emph{{DRS2022}}}. \bibinfo{address}{Bilbao}.
\newblock
\urldef\tempurl%
\url{https://doi.org/10.21606/drs.2022.266}
\showDOI{\tempurl}


\bibitem[Dew and Rosner(2019)]%
        {dew_designing_2019}
\bibfield{author}{\bibinfo{person}{Kristin~N. Dew} {and} \bibinfo{person}{Daniela~K. Rosner}.} \bibinfo{year}{2019}\natexlab{}.
\newblock \showarticletitle{Designing with {Waste}: {A} {Situated} {Inquiry} into the {Material} {Excess} of {Making}}. In \bibinfo{booktitle}{\emph{Proceedings of the 2019 on {Designing} {Interactive} {Systems} {Conference}}} \emph{(\bibinfo{series}{{DIS} '19})}. \bibinfo{publisher}{Association for Computing Machinery}, \bibinfo{address}{New York, NY, USA}, \bibinfo{pages}{1307--1319}.
\newblock
\showISBNx{978-1-4503-5850-7}
\urldef\tempurl%
\url{https://doi.org/10.1145/3322276.3322320}
\showDOI{\tempurl}


\bibitem[Dogan et~al\mbox{.}(2021)]%
        {dogan_sensicut_2021}
\bibfield{author}{\bibinfo{person}{Mustafa~Doga Dogan}, \bibinfo{person}{Steven~Vidal Acevedo~Colon}, \bibinfo{person}{Varnika Sinha}, \bibinfo{person}{Kaan Akşit}, {and} \bibinfo{person}{Stefanie Mueller}.} \bibinfo{year}{2021}\natexlab{}.
\newblock \showarticletitle{{SensiCut}: {Material}-{Aware} {Laser} {Cutting} {Using} {Speckle} {Sensing} and {Deep} {Learning}}. In \bibinfo{booktitle}{\emph{The 34th {Annual} {ACM} {Symposium} on {User} {Interface} {Software} and {Technology}}} \emph{(\bibinfo{series}{{UIST} '21})}. \bibinfo{publisher}{Association for Computing Machinery}, \bibinfo{address}{New York, NY, USA}, \bibinfo{pages}{24--38}.
\newblock
\showISBNx{978-1-4503-8635-7}
\urldef\tempurl%
\url{https://doi.org/10.1145/3472749.3474733}
\showDOI{\tempurl}


\bibitem[Dogan et~al\mbox{.}(2022)]%
        {dogan_fabricate_2022}
\bibfield{author}{\bibinfo{person}{Mustafa~Doga Dogan}, \bibinfo{person}{Patrick Baudisch}, \bibinfo{person}{Hrvoje Benko}, \bibinfo{person}{Michael Nebeling}, \bibinfo{person}{Huaishu Peng}, \bibinfo{person}{Valkyrie Savage}, {and} \bibinfo{person}{Stefanie Mueller}.} \bibinfo{year}{2022}\natexlab{}.
\newblock \showarticletitle{Fabricate {It} or {Render} {It}? {Digital} {Fabrication} vs. {Virtual} {Reality} for {Creating} {Objects} {Instantly}}. In \bibinfo{booktitle}{\emph{Extended {Abstracts} of the 2022 {CHI} {Conference} on {Human} {Factors} in {Computing} {Systems}}}. \bibinfo{publisher}{Association for Computing Machinery}, \bibinfo{pages}{5}.
\newblock
\urldef\tempurl%
\url{https://doi.org/10.1145/3491101.3516510}
\showDOI{\tempurl}


\bibitem[Efrat et~al\mbox{.}(2016)]%
        {efrat_hybrid_2016}
\bibfield{author}{\bibinfo{person}{Tamara~Anna Efrat}, \bibinfo{person}{Moran Mizrahi}, {and} \bibinfo{person}{Amit Zoran}.} \bibinfo{year}{2016}\natexlab{}.
\newblock \showarticletitle{The {Hybrid} {Bricolage}: {Bridging} {Parametric} {Design} with {Craft} through {Algorithmic} {Modularity}}. In \bibinfo{booktitle}{\emph{Proceedings of the 2016 {CHI} {Conference} on {Human} {Factors} in {Computing} {Systems}}}. \bibinfo{publisher}{ACM}, \bibinfo{address}{San Jose California USA}, \bibinfo{pages}{5984--5995}.
\newblock
\showISBNx{978-1-4503-3362-7}
\urldef\tempurl%
\url{https://doi.org/10.1145/2858036.2858441}
\showDOI{\tempurl}


\bibitem[Hoadley(2000)]%
        {hoadley_understanding_2000}
\bibfield{author}{\bibinfo{person}{R.~Bruce Hoadley}.} \bibinfo{year}{2000}\natexlab{}.
\newblock \bibinfo{booktitle}{\emph{Understanding {Wood}: {A} {Craftsman}'s {Guide} to {Wood} {Technology}}}.
\newblock \bibinfo{publisher}{Taunton Press}.
\newblock
\showISBNx{978-1-56158-358-4}


\bibitem[Htet~Kyaw et~al\mbox{.}(2024)]%
        {htet_kyaw_gesture_2024}
\bibfield{author}{\bibinfo{person}{Alexander Htet~Kyaw}, \bibinfo{person}{Lawson Spencer}, \bibinfo{person}{Sasa Zivkovic}, {and} \bibinfo{person}{Leslie Lok}.} \bibinfo{year}{2024}\natexlab{}.
\newblock \showarticletitle{Gesture {Recognition} for {Feedback} {Based} {Mixed} {Reality} and {Robotic} {Fabrication}: {A} {Case} {Study} of the {UnLog} {Tower}}. In \bibinfo{booktitle}{\emph{Phygital {Intelligence}}}, \bibfield{editor}{\bibinfo{person}{Chao Yan}, \bibinfo{person}{Hua Chai}, \bibinfo{person}{Tongyue Sun}, {and} \bibinfo{person}{Philip~F. Yuan}} (Eds.). \bibinfo{publisher}{Springer Nature}, \bibinfo{address}{Singapore}, \bibinfo{pages}{331--345}.
\newblock
\showISBNx{978-981-9984-05-3}
\urldef\tempurl%
\url{https://doi.org/10.1007/978-981-99-8405-3_28}
\showDOI{\tempurl}


\bibitem[Jahn et~al\mbox{.}(2024)]%
        {jahn_mixed_2024}
\bibfield{author}{\bibinfo{person}{Gwyllim Jahn}, \bibinfo{person}{Cameron Newnham}, {and} \bibinfo{person}{Nick Berg}.} \bibinfo{year}{2024}\natexlab{}.
\newblock \showarticletitle{Mixed {Reality} {Carpentry}}. In \bibinfo{booktitle}{\emph{{RobArch} 2024}}. \bibinfo{address}{Toronto}.
\newblock
\urldef\tempurl%
\url{https://www.researchgate.net/publication/381189449_Mixed_Reality_Carpentry}
\showURL{%
\tempurl}


\bibitem[Johns et~al\mbox{.}(2023)]%
        {johns_framework_2023}
\bibfield{author}{\bibinfo{person}{Ryan~Luke Johns}, \bibinfo{person}{Martin Wermelinger}, \bibinfo{person}{Ruben Mascaro}, \bibinfo{person}{Dominic Jud}, \bibinfo{person}{Ilmar Hurkxkens}, \bibinfo{person}{Lauren Vasey}, \bibinfo{person}{Margarita Chli}, \bibinfo{person}{Fabio Gramazio}, \bibinfo{person}{Matthias Kohler}, {and} \bibinfo{person}{Marco Hutter}.} \bibinfo{year}{2023}\natexlab{}.
\newblock \showarticletitle{A framework for robotic excavation and dry stone construction using on-site materials}.
\newblock \bibinfo{journal}{\emph{Science Robotics}} \bibinfo{volume}{8}, \bibinfo{number}{84} (\bibinfo{date}{Nov.} \bibinfo{year}{2023}), \bibinfo{pages}{eabp9758}.
\newblock
\urldef\tempurl%
\url{https://doi.org/10.1126/scirobotics.abp9758}
\showDOI{\tempurl}
\newblock
\shownote{Publisher: American Association for the Advancement of Science}.


\bibitem[Kim et~al\mbox{.}(2019)]%
        {kim_virtualcomponent_2019}
\bibfield{author}{\bibinfo{person}{Yoonji Kim}, \bibinfo{person}{Youngkyung Choi}, \bibinfo{person}{Hyein Lee}, \bibinfo{person}{Geehyuk Lee}, {and} \bibinfo{person}{Andrea Bianchi}.} \bibinfo{year}{2019}\natexlab{}.
\newblock \showarticletitle{{VirtualComponent}: {A} {Mixed}-{Reality} {Tool} for {Designing} and {Tuning} {Breadboarded} {Circuits}}. In \bibinfo{booktitle}{\emph{Proceedings of the 2019 {CHI} {Conference} on {Human} {Factors} in {Computing} {Systems}}}. \bibinfo{publisher}{ACM}, \bibinfo{address}{Glasgow Scotland Uk}, \bibinfo{pages}{1--13}.
\newblock
\showISBNx{978-1-4503-5970-2}
\urldef\tempurl%
\url{https://doi.org/10.1145/3290605.3300407}
\showDOI{\tempurl}


\bibitem[Koo et~al\mbox{.}(2017)]%
        {koo_towards_2017}
\bibfield{author}{\bibinfo{person}{Bongjin Koo}, \bibinfo{person}{Jean Hergel}, \bibinfo{person}{Sylvain Lefebvre}, {and} \bibinfo{person}{Niloy~J. Mitra}.} \bibinfo{year}{2017}\natexlab{}.
\newblock \showarticletitle{Towards {Zero}-{Waste} {Furniture} {Design}}.
\newblock \bibinfo{journal}{\emph{IEEE Transactions on Visualization and Computer Graphics}} \bibinfo{volume}{23}, \bibinfo{number}{12} (\bibinfo{date}{Dec.} \bibinfo{year}{2017}), \bibinfo{pages}{2627--2640}.
\newblock
\showISSN{1077-2626}
\urldef\tempurl%
\url{https://doi.org/10.1109/TVCG.2016.2633519}
\showDOI{\tempurl}


\bibitem[Kyaw(2023)]%
        {kyaw_active_2023}
\bibfield{author}{\bibinfo{person}{Alexander~Htet Kyaw}.} \bibinfo{year}{2023}\natexlab{}.
\newblock \showarticletitle{Active {Bending} in {Physics}-{Based} {Mixed} {Reality}: {The} design and fabrication of a reconfigurable modular bamboo system}. In \bibinfo{booktitle}{\emph{Dokonal, {W}, {Hirschberg}, {U} and {Wurzer}, {G} (eds.), {Digital} {Design} {Reconsidered} - {Proceedings} of the 41st {Conference} on {Education} and {Research} in {Computer} {Aided} {Architectural} {Design} in {Europe} ({eCAADe} 2023) - {Volume} 1, {Graz}, 20-22 {September} 2023, pp. 169–178}}. \bibinfo{publisher}{CUMINCAD}.
\newblock
\urldef\tempurl%
\url{https://papers.cumincad.org/cgi-bin/works/paper/ecaade2023_447}
\showURL{%
\tempurl}


\bibitem[Larsson et~al\mbox{.}(2019)]%
        {larsson_human---loop_2019}
\bibfield{author}{\bibinfo{person}{Maria Larsson}, \bibinfo{person}{Hironori Yoshida}, {and} \bibinfo{person}{Takeo Igarashi}.} \bibinfo{year}{2019}\natexlab{}.
\newblock \showarticletitle{Human-in-the-loop fabrication of {3D} surfaces with natural tree branches}. In \bibinfo{booktitle}{\emph{Proceedings of the {ACM} {Symposium} on {Computational} {Fabrication}}}. \bibinfo{publisher}{ACM}, \bibinfo{address}{Pittsburgh Pennsylvania}, \bibinfo{pages}{1--12}.
\newblock
\showISBNx{978-1-4503-6795-0}
\urldef\tempurl%
\url{https://doi.org/10.1145/3328939.3329000}
\showDOI{\tempurl}


\bibitem[Larsson et~al\mbox{.}(2020)]%
        {larsson_tsugite_2020}
\bibfield{author}{\bibinfo{person}{Maria Larsson}, \bibinfo{person}{Hironori Yoshida}, \bibinfo{person}{Nobuyuki Umetani}, {and} \bibinfo{person}{Takeo Igarashi}.} \bibinfo{year}{2020}\natexlab{}.
\newblock \showarticletitle{Tsugite: {Interactive} {Design} and {Fabrication} of {Wood} {Joints}}. In \bibinfo{booktitle}{\emph{Proceedings of the 33rd {Annual} {ACM} {Symposium} on {User} {Interface} {Software} and {Technology}}}. \bibinfo{publisher}{ACM}, \bibinfo{address}{Virtual Event USA}, \bibinfo{pages}{317--327}.
\newblock
\showISBNx{978-1-4503-7514-6}
\urldef\tempurl%
\url{https://doi.org/10.1145/3379337.3415899}
\showDOI{\tempurl}


\bibitem[Lau et~al\mbox{.}(2012)]%
        {lau_situated_2012}
\bibfield{author}{\bibinfo{person}{Manfred Lau}, \bibinfo{person}{Masaki Hirose}, \bibinfo{person}{Akira Ohgawara}, \bibinfo{person}{Jun Mitani}, {and} \bibinfo{person}{Takeo Igarashi}.} \bibinfo{year}{2012}\natexlab{}.
\newblock \showarticletitle{Situated modeling: a shape-stamping interface with tangible primitives}. In \bibinfo{booktitle}{\emph{Proceedings of the {Sixth} {International} {Conference} on {Tangible}, {Embedded} and {Embodied} {Interaction}}}. \bibinfo{publisher}{ACM}, \bibinfo{address}{Kingston Ontario Canada}, \bibinfo{pages}{275--282}.
\newblock
\showISBNx{978-1-4503-1174-8}
\urldef\tempurl%
\url{https://doi.org/10.1145/2148131.2148190}
\showDOI{\tempurl}


\bibitem[Lau et~al\mbox{.}(2011)]%
        {lau_converting_2011}
\bibfield{author}{\bibinfo{person}{Manfred Lau}, \bibinfo{person}{Akira Ohgawara}, \bibinfo{person}{Jun Mitani}, {and} \bibinfo{person}{Takeo Igarashi}.} \bibinfo{year}{2011}\natexlab{}.
\newblock \showarticletitle{Converting {3D} furniture models to fabricatable parts and connectors}. In \bibinfo{booktitle}{\emph{{ACM} {SIGGRAPH} 2011 papers}}. \bibinfo{publisher}{ACM}, \bibinfo{address}{Vancouver British Columbia Canada}, \bibinfo{pages}{1--6}.
\newblock
\showISBNx{978-1-4503-0943-1}
\urldef\tempurl%
\url{https://doi.org/10.1145/1964921.1964980}
\showDOI{\tempurl}


\bibitem[Lau et~al\mbox{.}(2010)]%
        {lau_modeling--context_2010}
\bibfield{author}{\bibinfo{person}{Manfred Lau}, \bibinfo{person}{Greg Saul}, \bibinfo{person}{Jun Mitani}, {and} \bibinfo{person}{Takeo Igarashi}.} \bibinfo{year}{2010}\natexlab{}.
\newblock \showarticletitle{Modeling-in-context: user design of complementary objects with a single photo}. In \bibinfo{booktitle}{\emph{Proceedings of the {Seventh} {Sketch}-{Based} {Interfaces} and {Modeling} {Symposium}}}. \bibinfo{publisher}{Eurographics Association}, \bibinfo{address}{Annecy, France}, \bibinfo{pages}{17--24}.
\newblock


\bibitem[Lee et~al\mbox{.}(2018)]%
        {lee_interactive_2018}
\bibfield{author}{\bibinfo{person}{Bokyung Lee}, \bibinfo{person}{Joongi Shin}, \bibinfo{person}{Hyoshin Bae}, {and} \bibinfo{person}{Daniel Saakes}.} \bibinfo{year}{2018}\natexlab{}.
\newblock \showarticletitle{Interactive and {Situated} {Guidelines} to {Help} {Users} {Design} a {Personal} {Desk} that {Fits} {Their} {Bodies}}. In \bibinfo{booktitle}{\emph{Proceedings of the 2018 {Designing} {Interactive} {Systems} {Conference}}}. \bibinfo{publisher}{ACM}, \bibinfo{address}{Hong Kong China}, \bibinfo{pages}{637--650}.
\newblock
\showISBNx{978-1-4503-5198-0}
\urldef\tempurl%
\url{https://doi.org/10.1145/3196709.3196725}
\showDOI{\tempurl}


\bibitem[Leen et~al\mbox{.}(2019)]%
        {leen_jigfab_2019}
\bibfield{author}{\bibinfo{person}{Danny Leen}, \bibinfo{person}{Tom Veuskens}, \bibinfo{person}{Kris Luyten}, {and} \bibinfo{person}{Raf Ramakers}.} \bibinfo{year}{2019}\natexlab{}.
\newblock \showarticletitle{{JigFab}: {Computational} {Fabrication} of {Constraints} to {Facilitate} {Woodworking} with {Power} {Tools}}. In \bibinfo{booktitle}{\emph{Proceedings of the 2019 {CHI} {Conference} on {Human} {Factors} in {Computing} {Systems}}} \emph{(\bibinfo{series}{{CHI} '19})}. \bibinfo{publisher}{Association for Computing Machinery}, \bibinfo{address}{New York, NY, USA}, \bibinfo{pages}{1--12}.
\newblock
\showISBNx{978-1-4503-5970-2}
\urldef\tempurl%
\url{https://doi.org/10.1145/3290605.3300386}
\showDOI{\tempurl}


\bibitem[Lin and Jo(2024)]%
        {lin_leatherboard_2024}
\bibfield{author}{\bibinfo{person}{Yujun Lin} {and} \bibinfo{person}{Jeyeon Jo}.} \bibinfo{year}{2024}\natexlab{}.
\newblock \showarticletitle{{LeatherBoard}: {Sustainable} {On}-body {Rapid} {Prototyping} with {Leather} {Scraps} and {Machine} {Embroidery}}. In \bibinfo{booktitle}{\emph{Extended {Abstracts} of the {CHI} {Conference} on {Human} {Factors} in {Computing} {Systems}}}. \bibinfo{publisher}{ACM}, \bibinfo{address}{Honolulu HI USA}, \bibinfo{pages}{1--7}.
\newblock
\showISBNx{9798400703317}
\urldef\tempurl%
\url{https://doi.org/10.1145/3613905.3650894}
\showDOI{\tempurl}


\bibitem[Lipton et~al\mbox{.}(2018)]%
        {lipton_robot_2018}
\bibfield{author}{\bibinfo{person}{Jeffrey~I Lipton}, \bibinfo{person}{Adriana Schulz}, \bibinfo{person}{Andrew Spielberg}, \bibinfo{person}{Luis Trueba}, \bibinfo{person}{Wojciech Matusik}, {and} \bibinfo{person}{Daniela Rus}.} \bibinfo{year}{2018}\natexlab{}.
\newblock \showarticletitle{Robot {Assisted} {Carpentry} for {Mass} {Customization}}. In \bibinfo{booktitle}{\emph{2018 {IEEE} {International} {Conference} on {Robotics} and {Automation} ({ICRA})}}. \bibinfo{pages}{3540--3547}.
\newblock
\urldef\tempurl%
\url{https://doi.org/10.1109/ICRA.2018.8460736}
\showDOI{\tempurl}
\newblock
\shownote{ISSN: 2577-087X}.


\bibitem[Lu et~al\mbox{.}(2023)]%
        {lu_ecoeda_2023}
\bibfield{author}{\bibinfo{person}{Jasmine Lu}, \bibinfo{person}{Beza Desta}, \bibinfo{person}{K.~D. Wu}, \bibinfo{person}{Romain Nith}, \bibinfo{person}{Joyce~E Passananti}, {and} \bibinfo{person}{Pedro Lopes}.} \bibinfo{year}{2023}\natexlab{}.
\newblock \showarticletitle{{ecoEDA}: {Recycling} {E}-waste {During} {Electronics} {Design}}. In \bibinfo{booktitle}{\emph{Proceedings of the 36th {Annual} {ACM} {Symposium} on {User} {Interface} {Software} and {Technology}}}. \bibinfo{publisher}{ACM}, \bibinfo{address}{San Francisco CA USA}, \bibinfo{pages}{1--14}.
\newblock
\showISBNx{9798400701320}
\urldef\tempurl%
\url{https://doi.org/10.1145/3586183.3606745}
\showDOI{\tempurl}


\bibitem[Magrisso et~al\mbox{.}(2018)]%
        {magrisso_digital_2018}
\bibfield{author}{\bibinfo{person}{Shiran Magrisso}, \bibinfo{person}{Moran Mizrahi}, {and} \bibinfo{person}{Amit Zoran}.} \bibinfo{year}{2018}\natexlab{}.
\newblock \showarticletitle{Digital {Joinery} {For} {Hybrid} {Carpentry}}. In \bibinfo{booktitle}{\emph{Proceedings of the 2018 {CHI} {Conference} on {Human} {Factors} in {Computing} {Systems}}} \emph{(\bibinfo{series}{{CHI} '18})}. \bibinfo{publisher}{Association for Computing Machinery}, \bibinfo{address}{New York, NY, USA}, \bibinfo{pages}{1--11}.
\newblock
\showISBNx{978-1-4503-5620-6}
\urldef\tempurl%
\url{https://doi.org/10.1145/3173574.3173741}
\showDOI{\tempurl}


\bibitem[Menges(2012)]%
        {menges_material_2012}
\bibfield{author}{\bibinfo{person}{Achim Menges}.} \bibinfo{year}{2012}\natexlab{}.
\newblock \showarticletitle{Material {Resourcefulness}: {Activating} {Material} {Information} in {Computational} {Design}}.
\newblock \bibinfo{journal}{\emph{Architectural Design}} \bibinfo{volume}{82}, \bibinfo{number}{2} (\bibinfo{year}{2012}), \bibinfo{pages}{34--43}.
\newblock
\showISSN{1554-2769}
\urldef\tempurl%
\url{https://doi.org/10.1002/ad.1377}
\showDOI{\tempurl}
\newblock
\shownote{\_eprint: https://onlinelibrary.wiley.com/doi/pdf/10.1002/ad.1377}.


\bibitem[Mitterberger et~al\mbox{.}(2022)]%
        {mitterberger_interactive_2022}
\bibfield{author}{\bibinfo{person}{Daniela Mitterberger}, \bibinfo{person}{Selen Ercan~Jenny}, \bibinfo{person}{Lauren Vasey}, \bibinfo{person}{Ena Lloret-Fritschi}, \bibinfo{person}{Petrus Aejmelaeus-Lindström}, \bibinfo{person}{Fabio Gramazio}, {and} \bibinfo{person}{Matthias Kohler}.} \bibinfo{year}{2022}\natexlab{}.
\newblock \showarticletitle{Interactive {Robotic} {Plastering}: {Augmented} {Interactive} {Design} and {Fabrication} for {On}-site {Robotic} {Plastering}}. In \bibinfo{booktitle}{\emph{{CHI} {Conference} on {Human} {Factors} in {Computing} {Systems}}}. \bibinfo{publisher}{ACM}, \bibinfo{address}{New Orleans LA USA}, \bibinfo{pages}{1--18}.
\newblock
\showISBNx{978-1-4503-9157-3}
\urldef\tempurl%
\url{https://doi.org/10.1145/3491102.3501842}
\showDOI{\tempurl}


\bibitem[{Mollica, Zachary} and {Self, Martin}(2016)]%
        {mollica_zachary_tree_2016}
\bibfield{author}{\bibinfo{person}{{Mollica, Zachary}} {and} \bibinfo{person}{{Self, Martin}}.} \bibinfo{year}{2016}\natexlab{}.
\newblock \bibinfo{booktitle}{\emph{Tree {Fork} {Truss}: {Geometric} {Strategies} for {Exploiting} {Inherent} {Material} {Form}}}.
\newblock \bibinfo{publisher}{Advances in Architectural Geometry}.
\newblock
\urldef\tempurl%
\url{https://doi.org/10.3218/3778-4_11}
\showDOI{\tempurl}


\bibitem[Nordmoen and McPherson(2022)]%
        {nordmoen_making_2022}
\bibfield{author}{\bibinfo{person}{Charlotte Nordmoen} {and} \bibinfo{person}{Andrew~P. McPherson}.} \bibinfo{year}{2022}\natexlab{}.
\newblock \showarticletitle{Making space for material entanglements: {A} diffractive analysis of woodwork and the practice of making an interactive system}. In \bibinfo{booktitle}{\emph{Proceedings of the 2022 {ACM} {Designing} {Interactive} {Systems} {Conference}}} \emph{(\bibinfo{series}{{DIS} '22})}. \bibinfo{publisher}{Association for Computing Machinery}, \bibinfo{address}{New York, NY, USA}, \bibinfo{pages}{415--423}.
\newblock
\showISBNx{978-1-4503-9358-4}
\urldef\tempurl%
\url{https://doi.org/10.1145/3532106.3533572}
\showDOI{\tempurl}


\bibitem[Nuernberger et~al\mbox{.}(2016)]%
        {nuernberger_snaptoreality_2016}
\bibfield{author}{\bibinfo{person}{Benjamin Nuernberger}, \bibinfo{person}{Eyal Ofek}, \bibinfo{person}{Hrvoje Benko}, {and} \bibinfo{person}{Andrew~D. Wilson}.} \bibinfo{year}{2016}\natexlab{}.
\newblock \showarticletitle{{SnapToReality}: {Aligning} {Augmented} {Reality} to the {Real} {World}}. In \bibinfo{booktitle}{\emph{Proceedings of the 2016 {CHI} {Conference} on {Human} {Factors} in {Computing} {Systems}}}. \bibinfo{publisher}{ACM}, \bibinfo{address}{San Jose California USA}, \bibinfo{pages}{1233--1244}.
\newblock
\showISBNx{978-1-4503-3362-7}
\urldef\tempurl%
\url{https://doi.org/10.1145/2858036.2858250}
\showDOI{\tempurl}


\bibitem[Parry and Guy(2021)]%
        {parry_recycling_2021}
\bibfield{author}{\bibinfo{person}{Caitlyn Parry} {and} \bibinfo{person}{Sean Guy}.} \bibinfo{year}{2021}\natexlab{}.
\newblock \showarticletitle{Recycling {Construction} {Waste} {Material} with the {Use} of {AR}}.
\newblock In \bibinfo{booktitle}{\emph{2020 {DigitalFUTURES}}}. \bibinfo{publisher}{Springer}, \bibinfo{pages}{57--67}.
\newblock
\showISBNx{978-981-334-399-3}
\urldef\tempurl%
\url{https://doi.org/10.1007/978-981-33-4400-6_6}
\showDOI{\tempurl}


\bibitem[Ramakers et~al\mbox{.}(2023)]%
        {ramakers_measurement_2023}
\bibfield{author}{\bibinfo{person}{Raf Ramakers}, \bibinfo{person}{Danny Leen}, \bibinfo{person}{Jeeeun Kim}, \bibinfo{person}{Kris Luyten}, \bibinfo{person}{Steven Houben}, {and} \bibinfo{person}{Tom Veuskens}.} \bibinfo{year}{2023}\natexlab{}.
\newblock \showarticletitle{Measurement {Patterns}: {User}-{Oriented} {Strategies} for {Dealing} with {Measurements} and {Dimensions} in {Making} {Processes}}. In \bibinfo{booktitle}{\emph{Proceedings of the 2023 {CHI} {Conference} on {Human} {Factors} in {Computing} {Systems}}}. \bibinfo{publisher}{ACM}, \bibinfo{address}{Hamburg Germany}, \bibinfo{pages}{1--17}.
\newblock
\showISBNx{978-1-4503-9421-5}
\urldef\tempurl%
\url{https://doi.org/10.1145/3544548.3581157}
\showDOI{\tempurl}


\bibitem[Reipschläger and Dachselt(2019)]%
        {reipschlager_designar_2019}
\bibfield{author}{\bibinfo{person}{Patrick Reipschläger} {and} \bibinfo{person}{Raimund Dachselt}.} \bibinfo{year}{2019}\natexlab{}.
\newblock \showarticletitle{{DesignAR}: {Immersive} {3D}-{Modeling} {Combining} {Augmented} {Reality} with {Interactive} {Displays}}. In \bibinfo{booktitle}{\emph{Proceedings of the 2019 {ACM} {International} {Conference} on {Interactive} {Surfaces} and {Spaces}}}. \bibinfo{publisher}{ACM}, \bibinfo{address}{Daejeon Republic of Korea}, \bibinfo{pages}{29--41}.
\newblock
\showISBNx{978-1-4503-6891-9}
\urldef\tempurl%
\url{https://doi.org/10.1145/3343055.3359718}
\showDOI{\tempurl}


\bibitem[Saakes et~al\mbox{.}(2013)]%
        {saakes_paccam_2013}
\bibfield{author}{\bibinfo{person}{Daniel Saakes}, \bibinfo{person}{Thomas Cambazard}, \bibinfo{person}{Jun Mitani}, {and} \bibinfo{person}{Takeo Igarashi}.} \bibinfo{year}{2013}\natexlab{}.
\newblock \showarticletitle{{PacCAM}: material capture and interactive {2D} packing for efficient material usage on {CNC} cutting machines}. In \bibinfo{booktitle}{\emph{Proceedings of the 26th {Annual} {ACM} {Symposium} on {User} {Interface} {Software} and {Technology}}} \emph{(\bibinfo{series}{{UIST} '13})}. \bibinfo{publisher}{Association for Computing Machinery}, \bibinfo{address}{New York, NY, USA}, \bibinfo{pages}{441--446}.
\newblock
\showISBNx{978-1-4503-2268-3}
\urldef\tempurl%
\url{https://doi.org/10.1145/2501988.2501990}
\showDOI{\tempurl}
\newblock
\shownote{event-place: St. Andrews, Scotland, United Kingdom}.


\bibitem[Saul et~al\mbox{.}(2010)]%
        {saul_sketchchair_2010}
\bibfield{author}{\bibinfo{person}{Greg Saul}, \bibinfo{person}{Manfred Lau}, \bibinfo{person}{Jun Mitani}, {and} \bibinfo{person}{Takeo Igarashi}.} \bibinfo{year}{2010}\natexlab{}.
\newblock \showarticletitle{{SketchChair}: an all-in-one chair design system for end users}. In \bibinfo{booktitle}{\emph{Proceedings of the fifth international conference on {Tangible}, embedded, and embodied interaction}}. \bibinfo{publisher}{ACM}, \bibinfo{address}{Funchal Portugal}, \bibinfo{pages}{73--80}.
\newblock
\showISBNx{978-1-4503-0478-8}
\urldef\tempurl%
\url{https://doi.org/10.1145/1935701.1935717}
\showDOI{\tempurl}


\bibitem[Schoop et~al\mbox{.}(2016)]%
        {schoop_drill_2016}
\bibfield{author}{\bibinfo{person}{Eldon Schoop}, \bibinfo{person}{Michelle Nguyen}, \bibinfo{person}{Daniel Lim}, \bibinfo{person}{Valkyrie Savage}, \bibinfo{person}{Sean Follmer}, {and} \bibinfo{person}{Björn Hartmann}.} \bibinfo{year}{2016}\natexlab{}.
\newblock \showarticletitle{Drill {Sergeant}: {Supporting} {Physical} {Construction} {Projects} through an {Ecosystem} of {Augmented} {Tools}}. In \bibinfo{booktitle}{\emph{Proceedings of the 2016 {CHI} {Conference} {Extended} {Abstracts} on {Human} {Factors} in {Computing} {Systems}}}. \bibinfo{publisher}{ACM}, \bibinfo{address}{San Jose California USA}, \bibinfo{pages}{1607--1614}.
\newblock
\showISBNx{978-1-4503-4082-3}
\urldef\tempurl%
\url{https://doi.org/10.1145/2851581.2892429}
\showDOI{\tempurl}


\bibitem[Schulz et~al\mbox{.}(2014)]%
        {schulz_design_2014}
\bibfield{author}{\bibinfo{person}{Adriana Schulz}, \bibinfo{person}{Ariel Shamir}, \bibinfo{person}{David I.~W. Levin}, \bibinfo{person}{Pitchaya Sitthi-amorn}, {and} \bibinfo{person}{Wojciech Matusik}.} \bibinfo{year}{2014}\natexlab{}.
\newblock \showarticletitle{Design and fabrication by example}.
\newblock \bibinfo{journal}{\emph{ACM Transactions on Graphics}} \bibinfo{volume}{33}, \bibinfo{number}{4} (\bibinfo{date}{July} \bibinfo{year}{2014}), \bibinfo{pages}{1--11}.
\newblock
\showISSN{0730-0301, 1557-7368}
\urldef\tempurl%
\url{https://doi.org/10.1145/2601097.2601127}
\showDOI{\tempurl}


\bibitem[Sethapakdi et~al\mbox{.}(2021)]%
        {sethapakdi_fabricaide_2021}
\bibfield{author}{\bibinfo{person}{Ticha Sethapakdi}, \bibinfo{person}{Daniel Anderson}, \bibinfo{person}{Adrian Reginald~Chua Sy}, {and} \bibinfo{person}{Stefanie Mueller}.} \bibinfo{year}{2021}\natexlab{}.
\newblock \showarticletitle{Fabricaide: {Fabrication}-{Aware} {Design} for {2D} {Cutting} {Machines}}. In \bibinfo{booktitle}{\emph{Proceedings of the 2021 {CHI} {Conference} on {Human} {Factors} in {Computing} {Systems}}} \emph{(\bibinfo{series}{{CHI} '21})}. \bibinfo{publisher}{Association for Computing Machinery}, \bibinfo{address}{New York, NY, USA}, \bibinfo{pages}{1--12}.
\newblock
\showISBNx{978-1-4503-8096-6}
\urldef\tempurl%
\url{https://doi.org/10.1145/3411764.3445345}
\showDOI{\tempurl}


\bibitem[Shao et~al\mbox{.}(2016)]%
        {shao_dynamic_2016}
\bibfield{author}{\bibinfo{person}{Tianjia Shao}, \bibinfo{person}{Dongping Li}, \bibinfo{person}{Yuliang Rong}, \bibinfo{person}{Changxi Zheng}, {and} \bibinfo{person}{Kun Zhou}.} \bibinfo{year}{2016}\natexlab{}.
\newblock \showarticletitle{Dynamic furniture modeling through assembly instructions}.
\newblock \bibinfo{journal}{\emph{ACM Transactions on Graphics}} \bibinfo{volume}{35}, \bibinfo{number}{6} (\bibinfo{date}{Nov.} \bibinfo{year}{2016}), \bibinfo{pages}{1--15}.
\newblock
\showISSN{0730-0301, 1557-7368}
\urldef\tempurl%
\url{https://doi.org/10.1145/2980179.2982416}
\showDOI{\tempurl}


\bibitem[Stemasov et~al\mbox{.}(2022)]%
        {stemasov_ephemeral_2022}
\bibfield{author}{\bibinfo{person}{Evgeny Stemasov}, \bibinfo{person}{Alexander Botner}, \bibinfo{person}{Enrico Rukzio}, {and} \bibinfo{person}{Jan Gugenheimer}.} \bibinfo{year}{2022}\natexlab{}.
\newblock \showarticletitle{Ephemeral {Fabrication}: {Exploring} a {Ubiquitous} {Fabrication} {Scenario} of {Low}-{Effort}, {In}-{Situ} {Creation} of {Short}-{Lived} {Physical} {Artifacts}}. In \bibinfo{booktitle}{\emph{Sixteenth {International} {Conference} on {Tangible}, {Embedded}, and {Embodied} {Interaction}}}. \bibinfo{publisher}{ACM}, \bibinfo{address}{Daejeon Republic of Korea}, \bibinfo{pages}{1--17}.
\newblock
\showISBNx{978-1-4503-9147-4}
\urldef\tempurl%
\url{https://doi.org/10.1145/3490149.3501331}
\showDOI{\tempurl}


\bibitem[Stemasov et~al\mbox{.}(2024)]%
        {stemasov_param_2024}
\bibfield{author}{\bibinfo{person}{Evgeny Stemasov}, \bibinfo{person}{Simon Demharter}, \bibinfo{person}{Max Rädler}, \bibinfo{person}{Jan Gugenheimer}, {and} \bibinfo{person}{Enrico Rukzio}.} \bibinfo{year}{2024}\natexlab{}.
\newblock \showarticletitle{{pARam}: {Leveraging} {Parametric} {Design} in {Extended} {Reality} to {Support} the {Personalization} of {Artifacts} for {Personal} {Fabrication}}. In \bibinfo{booktitle}{\emph{Proceedings of the {CHI} {Conference} on {Human} {Factors} in {Computing} {Systems}}}. \bibinfo{publisher}{ACM}, \bibinfo{address}{Honolulu HI USA}, \bibinfo{pages}{1--22}.
\newblock
\showISBNx{9798400703300}
\urldef\tempurl%
\url{https://doi.org/10.1145/3613904.3642083}
\showDOI{\tempurl}


\bibitem[Stemasov et~al\mbox{.}(2023)]%
        {stemasov_brickstart_2023}
\bibfield{author}{\bibinfo{person}{Evgeny Stemasov}, \bibinfo{person}{Jessica Hohn}, \bibinfo{person}{Maurice Cordts}, \bibinfo{person}{Anja Schikorr}, \bibinfo{person}{Enrico Rukzio}, {and} \bibinfo{person}{Jan Gugenheimer}.} \bibinfo{year}{2023}\natexlab{}.
\newblock \showarticletitle{{BrickStARt}: {Enabling} {In}-situ {Design} and {Tangible} {Exploration} for {Personal} {Fabrication} using {Mixed} {Reality}}.
\newblock \bibinfo{journal}{\emph{Proceedings of the ACM on Human-Computer Interaction}} \bibinfo{volume}{7}, \bibinfo{number}{ISS} (\bibinfo{date}{Oct.} \bibinfo{year}{2023}), \bibinfo{pages}{64--92}.
\newblock
\showISSN{2573-0142}
\urldef\tempurl%
\url{https://doi.org/10.1145/3626465}
\showDOI{\tempurl}


\bibitem[Stemasov et~al\mbox{.}(2020)]%
        {stemasov_mixmatch_2020}
\bibfield{author}{\bibinfo{person}{Evgeny Stemasov}, \bibinfo{person}{Tobias Wagner}, \bibinfo{person}{Jan Gugenheimer}, {and} \bibinfo{person}{Enrico Rukzio}.} \bibinfo{year}{2020}\natexlab{}.
\newblock \showarticletitle{Mix\&{Match}: {Towards} {Omitting} {Modelling} {Through} {In}-situ {Remixing} of {Model} {Repository} {Artifacts} in {Mixed} {Reality}}. In \bibinfo{booktitle}{\emph{Proceedings of the 2020 {CHI} {Conference} on {Human} {Factors} in {Computing} {Systems}}}. \bibinfo{publisher}{ACM}, \bibinfo{address}{Honolulu HI USA}, \bibinfo{pages}{1--12}.
\newblock
\showISBNx{978-1-4503-6708-0}
\urldef\tempurl%
\url{https://doi.org/10.1145/3313831.3376839}
\showDOI{\tempurl}


\bibitem[Strandholt et~al\mbox{.}(2020)]%
        {strandholt_knock_2020}
\bibfield{author}{\bibinfo{person}{Patrick~L. Strandholt}, \bibinfo{person}{Oana~A. Dogaru}, \bibinfo{person}{Niels~C. Nilsson}, \bibinfo{person}{Rolf Nordahl}, {and} \bibinfo{person}{Stefania Serafin}.} \bibinfo{year}{2020}\natexlab{}.
\newblock \showarticletitle{Knock on {Wood}: {Combining} {Redirected} {Touching} and {Physical} {Props} for {Tool}-{Based} {Interaction} in {Virtual} {Reality}}. In \bibinfo{booktitle}{\emph{Proceedings of the 2020 {CHI} {Conference} on {Human} {Factors} in {Computing} {Systems}}} \emph{(\bibinfo{series}{{CHI} '20})}. \bibinfo{publisher}{Association for Computing Machinery}, \bibinfo{address}{New York, NY, USA}, \bibinfo{pages}{1--13}.
\newblock
\showISBNx{978-1-4503-6708-0}
\urldef\tempurl%
\url{https://doi.org/10.1145/3313831.3376303}
\showDOI{\tempurl}


\bibitem[Sunshine(2022)]%
        {sunshine_inventory_2022}
\bibfield{author}{\bibinfo{person}{Gil Sunshine}.} \bibinfo{year}{2022}\natexlab{}.
\newblock \bibinfo{title}{Inventory: {CAD} for {Medium} {Resolution} {Materials}}.
\newblock
\newblock
\urldef\tempurl%
\url{https://gilsunshine.com}
\showURL{%
\tempurl}
\newblock
\shownote{Master of Architecture Thesis}.


\bibitem[Tian and Paulos(2021)]%
        {tian_adroid_2021}
\bibfield{author}{\bibinfo{person}{Rundong Tian} {and} \bibinfo{person}{Eric Paulos}.} \bibinfo{year}{2021}\natexlab{}.
\newblock \showarticletitle{Adroid: {Augmenting} {Hands}-on {Making} with a {Collaborative} {Robot}}. In \bibinfo{booktitle}{\emph{The 34th {Annual} {ACM} {Symposium} on {User} {Interface} {Software} and {Technology}}}. \bibinfo{publisher}{ACM}, \bibinfo{address}{Virtual Event USA}, \bibinfo{pages}{270--281}.
\newblock
\showISBNx{978-1-4503-8635-7}
\urldef\tempurl%
\url{https://doi.org/10.1145/3472749.3474749}
\showDOI{\tempurl}


\bibitem[Tian et~al\mbox{.}(2018)]%
        {tian_matchsticks_2018}
\bibfield{author}{\bibinfo{person}{Rundong Tian}, \bibinfo{person}{Sarah Sterman}, \bibinfo{person}{Ethan Chiou}, \bibinfo{person}{Jeremy Warner}, {and} \bibinfo{person}{Eric Paulos}.} \bibinfo{year}{2018}\natexlab{}.
\newblock \showarticletitle{{MatchSticks}: {Woodworking} through {Improvisational} {Digital} {Fabrication}}. In \bibinfo{booktitle}{\emph{Proceedings of the 2018 {CHI} {Conference} on {Human} {Factors} in {Computing} {Systems}}} \emph{(\bibinfo{series}{{CHI} '18})}. \bibinfo{publisher}{Association for Computing Machinery}, \bibinfo{address}{New York, NY, USA}, \bibinfo{pages}{1--12}.
\newblock
\showISBNx{978-1-4503-5620-6}
\urldef\tempurl%
\url{https://doi.org/10.1145/3173574.3173723}
\showDOI{\tempurl}


\bibitem[Toka et~al\mbox{.}(2023)]%
        {toka_adaptable_2023}
\bibfield{author}{\bibinfo{person}{Mert Toka}, \bibinfo{person}{Samuelle Bourgault}, \bibinfo{person}{Camila Friedman-Gerlicz}, {and} \bibinfo{person}{Jennifer Jacobs}.} \bibinfo{year}{2023}\natexlab{}.
\newblock \showarticletitle{An {Adaptable} {Workflow} for {Manual}-{Computational} {Ceramic} {Surface} {Ornamentation}}. In \bibinfo{booktitle}{\emph{Proceedings of the 36th {Annual} {ACM} {Symposium} on {User} {Interface} {Software} and {Technology}}} \emph{(\bibinfo{series}{{UIST} '23})}. \bibinfo{publisher}{Association for Computing Machinery}, \bibinfo{address}{New York, NY, USA}, \bibinfo{pages}{1--15}.
\newblock
\showISBNx{9798400701320}
\urldef\tempurl%
\url{https://doi.org/10.1145/3586183.3606726}
\showDOI{\tempurl}


\bibitem[Tran~O'Leary et~al\mbox{.}(2024)]%
        {tran_oleary_tandem_2024}
\bibfield{author}{\bibinfo{person}{Jasper Tran~O'Leary}, \bibinfo{person}{Thrisha Ramesh}, \bibinfo{person}{Octi Zhang}, {and} \bibinfo{person}{Nadya Peek}.} \bibinfo{year}{2024}\natexlab{}.
\newblock \showarticletitle{Tandem: {Reproducible} {Digital} {Fabrication} {Workflows} as {Multimodal} {Programs}}. In \bibinfo{booktitle}{\emph{Proceedings of the {CHI} {Conference} on {Human} {Factors} in {Computing} {Systems}}}. \bibinfo{publisher}{ACM}, \bibinfo{address}{Honolulu HI USA}, \bibinfo{pages}{1--16}.
\newblock
\showISBNx{9798400703300}
\urldef\tempurl%
\url{https://doi.org/10.1145/3613904.3642751}
\showDOI{\tempurl}


\bibitem[Umetani et~al\mbox{.}(2012)]%
        {umetani_guided_2012}
\bibfield{author}{\bibinfo{person}{Nobuyuki Umetani}, \bibinfo{person}{Takeo Igarashi}, {and} \bibinfo{person}{Niloy~J. Mitra}.} \bibinfo{year}{2012}\natexlab{}.
\newblock \showarticletitle{Guided exploration of physically valid shapes for furniture design}.
\newblock \bibinfo{journal}{\emph{ACM Transactions on Graphics}} \bibinfo{volume}{31}, \bibinfo{number}{4} (\bibinfo{date}{Aug.} \bibinfo{year}{2012}), \bibinfo{pages}{1--11}.
\newblock
\showISSN{0730-0301, 1557-7368}
\urldef\tempurl%
\url{https://doi.org/10.1145/2185520.2185582}
\showDOI{\tempurl}


\bibitem[Wall et~al\mbox{.}(2021)]%
        {wall_scrappy_2021}
\bibfield{author}{\bibinfo{person}{Ludwig~Wilhelm Wall}, \bibinfo{person}{Alec Jacobson}, \bibinfo{person}{Daniel Vogel}, {and} \bibinfo{person}{Oliver Schneider}.} \bibinfo{year}{2021}\natexlab{}.
\newblock \showarticletitle{Scrappy: {Using} {Scrap} {Material} as {Infill} to {Make} {Fabrication} {More} {Sustainable}}. In \bibinfo{booktitle}{\emph{Proceedings of the 2021 {CHI} {Conference} on {Human} {Factors} in {Computing} {Systems}}}. \bibinfo{publisher}{ACM}, \bibinfo{address}{Yokohama Japan}, \bibinfo{pages}{1--12}.
\newblock
\showISBNx{978-1-4503-8096-6}
\urldef\tempurl%
\url{https://doi.org/10.1145/3411764.3445187}
\showDOI{\tempurl}


\bibitem[Wall et~al\mbox{.}(2023)]%
        {wall_substiports_2023}
\bibfield{author}{\bibinfo{person}{Ludwig~Wilhelm Wall}, \bibinfo{person}{Oliver Schneider}, {and} \bibinfo{person}{Daniel Vogel}.} \bibinfo{year}{2023}\natexlab{}.
\newblock \showarticletitle{Substiports: {User}-{Inserted} {Ad} {Hoc} {Objects} as {Reusable} {Structural} {Support} for {Unmodified} {FDM} {3D} {Printers}}. In \bibinfo{booktitle}{\emph{Proceedings of the 36th {Annual} {ACM} {Symposium} on {User} {Interface} {Software} and {Technology}}}. \bibinfo{publisher}{ACM}, \bibinfo{address}{San Francisco CA USA}, \bibinfo{pages}{1--20}.
\newblock
\showISBNx{9798400701320}
\urldef\tempurl%
\url{https://doi.org/10.1145/3586183.3606718}
\showDOI{\tempurl}


\bibitem[Weichel et~al\mbox{.}(2015)]%
        {weichel_spata_2015}
\bibfield{author}{\bibinfo{person}{Christian Weichel}, \bibinfo{person}{Jason Alexander}, \bibinfo{person}{Abhijit Karnik}, {and} \bibinfo{person}{Hans Gellersen}.} \bibinfo{year}{2015}\natexlab{}.
\newblock \showarticletitle{{SPATA}: {Spatio}-{Tangible} {Tools} for {Fabrication}-{Aware} {Design}}. In \bibinfo{booktitle}{\emph{Proceedings of the {Ninth} {International} {Conference} on {Tangible}, {Embedded}, and {Embodied} {Interaction}}}. \bibinfo{publisher}{ACM}, \bibinfo{address}{Stanford California USA}, \bibinfo{pages}{189--196}.
\newblock
\showISBNx{978-1-4503-3305-4}
\urldef\tempurl%
\url{https://doi.org/10.1145/2677199.2680576}
\showDOI{\tempurl}


\bibitem[Weichel et~al\mbox{.}(2014)]%
        {weichel_mixfab_2014}
\bibfield{author}{\bibinfo{person}{Christian Weichel}, \bibinfo{person}{Manfred Lau}, \bibinfo{person}{David Kim}, \bibinfo{person}{Nicolas Villar}, {and} \bibinfo{person}{Hans~W. Gellersen}.} \bibinfo{year}{2014}\natexlab{}.
\newblock \showarticletitle{{MixFab}: a mixed-reality environment for personal fabrication}. In \bibinfo{booktitle}{\emph{Proceedings of the {SIGCHI} {Conference} on {Human} {Factors} in {Computing} {Systems}}}. \bibinfo{publisher}{ACM}, \bibinfo{address}{Toronto Ontario Canada}, \bibinfo{pages}{3855--3864}.
\newblock
\showISBNx{978-1-4503-2473-1}
\urldef\tempurl%
\url{https://doi.org/10.1145/2556288.2557090}
\showDOI{\tempurl}


\bibitem[Wu et~al\mbox{.}(2019)]%
        {wu_carpentry_2019}
\bibfield{author}{\bibinfo{person}{Chenming Wu}, \bibinfo{person}{Haisen Zhao}, \bibinfo{person}{Chandrakana Nandi}, \bibinfo{person}{Jeffrey~I. Lipton}, \bibinfo{person}{Zachary Tatlock}, {and} \bibinfo{person}{Adriana Schulz}.} \bibinfo{year}{2019}\natexlab{}.
\newblock \showarticletitle{Carpentry compiler}.
\newblock \bibinfo{journal}{\emph{ACM Trans. Graph.}} \bibinfo{volume}{38}, \bibinfo{number}{6} (\bibinfo{date}{Nov.} \bibinfo{year}{2019}), \bibinfo{pages}{Article No. 195}.
\newblock
\showISSN{0730-0301}
\urldef\tempurl%
\url{https://doi.org/10.1145/3355089.3356518}
\showDOI{\tempurl}
\newblock
\shownote{Place: New York, NY, USA Publisher: Association for Computing Machinery}.


\bibitem[Yanpanyanon et~al\mbox{.}(2020)]%
        {yanpanyanon_spatial_2020}
\bibfield{author}{\bibinfo{person}{Sakeson Yanpanyanon}, \bibinfo{person}{Peng Jiang}, {and} \bibinfo{person}{Takamitsu Tanaka}.} \bibinfo{year}{2020}\natexlab{}.
\newblock \showarticletitle{{SPATIAL} {REASONING} {ENHANCEMENT} {FOR} {DIY} {FURNITURE} {ASSEMBLY} {USING} {THE} {AUGMENTED} {REALITY} {APPLICATION}}.
\newblock \bibinfo{journal}{\emph{Journal of the Science of Design}} \bibinfo{volume}{4}, \bibinfo{number}{2} (\bibinfo{year}{2020}), \bibinfo{pages}{103--112}.
\newblock


\bibitem[Yao et~al\mbox{.}(2017)]%
        {yao_interactive_2017}
\bibfield{author}{\bibinfo{person}{Jiaxian Yao}, \bibinfo{person}{Danny~M. Kaufman}, \bibinfo{person}{Yotam Gingold}, {and} \bibinfo{person}{Maneesh Agrawala}.} \bibinfo{year}{2017}\natexlab{}.
\newblock \showarticletitle{Interactive {Design} and {Stability} {Analysis} of {Decorative} {Joinery} for {Furniture}}.
\newblock \bibinfo{journal}{\emph{ACM Transactions on Graphics}} \bibinfo{volume}{36}, \bibinfo{number}{2} (\bibinfo{date}{April} \bibinfo{year}{2017}), \bibinfo{pages}{1--16}.
\newblock
\showISSN{0730-0301, 1557-7368}
\urldef\tempurl%
\url{https://doi.org/10.1145/3054740}
\showDOI{\tempurl}


\bibitem[Zhao et~al\mbox{.}(2022)]%
        {zhao_co-optimization_2022}
\bibfield{author}{\bibinfo{person}{Haisen Zhao}, \bibinfo{person}{Max Willsey}, \bibinfo{person}{Amy Zhu}, \bibinfo{person}{Chandrakana Nandi}, \bibinfo{person}{Zachary Tatlock}, \bibinfo{person}{Justin Solomon}, {and} \bibinfo{person}{Adriana Schulz}.} \bibinfo{year}{2022}\natexlab{}.
\newblock \showarticletitle{Co-{Optimization} of {Design} and {Fabrication} {Plans} for {Carpentry}}.
\newblock \bibinfo{journal}{\emph{ACM Transactions on Graphics}} \bibinfo{volume}{41}, \bibinfo{number}{3} (\bibinfo{date}{March} \bibinfo{year}{2022}), \bibinfo{pages}{32:1--32:13}.
\newblock
\showISSN{0730-0301}
\urldef\tempurl%
\url{https://doi.org/10.1145/3508499}
\showDOI{\tempurl}


\bibitem[Zoran et~al\mbox{.}(2013)]%
        {zoran_human-computer_2013}
\bibfield{author}{\bibinfo{person}{Amit Zoran}, \bibinfo{person}{Roy Shilkrot}, {and} \bibinfo{person}{Joseph Paradiso}.} \bibinfo{year}{2013}\natexlab{}.
\newblock \showarticletitle{Human-computer interaction for hybrid carving}. In \bibinfo{booktitle}{\emph{Proceedings of the 26th annual {ACM} symposium on {User} interface software and technology}}. \bibinfo{publisher}{ACM}, \bibinfo{address}{St. Andrews Scotland, United Kingdom}, \bibinfo{pages}{433--440}.
\newblock
\showISBNx{978-1-4503-2268-3}
\urldef\tempurl%
\url{https://doi.org/10.1145/2501988.2502023}
\showDOI{\tempurl}


\end{thebibliography}

\end{document}